\documentclass[smallextended]{svjour3}       
\smartqed  

\usepackage{graphicx,psfrag,color}
\usepackage{dcolumn}
\usepackage{bm}
\usepackage{mathbbol}
\usepackage{amssymb,amsmath,amsfonts,amssymb}

\newcommand{\ket}[1]{\left|#1\right\rangle}      
\newcommand{\bra}[1]{\left\langle #1\right|}     

\begin{document}

\title{Asymptotic security analysis of teleportation based quantum cryptography}
\titlerunning{Security analysis of teleportation based quantum cryptography}

\author{Diogo Lima \and Gustavo Rigolin}

\institute{Diogo Lima \at
           Departamento de F\'{i}sica,
	   Universidade Federal de S\~ao Carlos, S\~ao Carlos, SP 13565-905,
	   Brazil          
          \and
           Gustavo Rigolin \at
           Departamento de F\'{i}sica,
	   Universidade Federal de S\~ao Carlos, S\~ao Carlos, SP 13565-905,
	   Brazil \\  
           \email{rigolin@ufscar.br}  
}

\date{Received: date / Accepted: date}

\maketitle

\begin{abstract}
We prove that the teleportation based quantum cryptography protocol presented in
[Opt. Commun. \textbf{283}, 184 (2010)], which is built using 
only orthogonal states encoding the classical bits that are teleported from Alice to Bob,
is asymptotically secure against all types of individual and collective attacks. We then investigate 
modifications to that protocol leading to greater secret-key rates and to 
security against coherent attacks. In other words, we show 
an unconditional secure quantum key distribution
protocol that does not need non-orthogonal quantum states to encode the bits of the secret key sent
from Alice to Bob. We also revisit the security proof of the BB84 protocol by exploring the 
non-uniqueness of the Schmidt decomposition of its entanglement-based representation. This allows us
to arrive at a secure transmission of the key for a slightly greater quantum bit error rate 
(quantum communication channel's noise) when compared to its standard security analysis. 
\end{abstract}


\maketitle

\section{Introduction} 

In today's world, where the amount of information produced, stored, and transmitted 
has increased to an unprecedented level, it is of utmost importance the
construction of ways to store and transmit information in a secure way, where only 
authorized parties are able to access the content of the stored and transmitted data \cite{sin00}.
Cryptography is an interdisciplinary field of research whose main goal is the construction 
of devices and protocols such that the transmission of information can be made in a secure
way, preventing an eavesdropper (Eve) from deciphering the message sent from one party (Alice)
to another one (Bob).

The standard solution currently employed by Alice to cipher a message, such that only Bob can decipher it,
is based on public key protocols \cite{riv79}. The security of those protocols is based on the assumption that 
there is no efficient algorithm to factor huge prime numbers. If this were the case, this algorithm could be
adapted to break the security of all public key cryptography protocols. Although
there is no classical algorithm that can efficiently factor prime numbers of any size, this is not the
case quantumly \cite{shor97}. The proof that a quantum computer can factor efficiently large prime 
numbers was the main trigger that boosted the field of quantum cryptography, which had already offered an alternative
solution to secure communication  \cite{ben84} about 10 years before Peter Shor's work \cite{shor97}. 

The protocol presented by Bennett and Brassard in 1984 \cite{ben84}, 
later called the BB84 protocol, was the quantum solution
to the key distribution problem. The key distribution problem is basically the classical impossibility to be $100\%$ sure
that only two parties, and no one else, agree on a random sequence of bits that is sent from one party to another.
The transmission of the key from Alice to Bob via classical means 
can in principle be monitored by a clever and powerful enough Eve, who
copies the random bits during their transmission without ever being detected by Alice and Bob. This is the scenario 
dictated by classical physics, in which copying or cloning of bits is always possible. 
However, if we use non-orthogonal quantum states to encode the bits to be transmitted from Alice to Bob, and if we
believe in the correctness of quantum mechanics, no
eavesdropper can tamper with the quantum key distribution without being discovered by Alice and Bob \cite{ben84}.
The quantum solution to the key distribution problem brought back to the table classical private key protocols,
the one-time pad for example \cite{sin00}, which can now be made secure if the key distribution needed 
to its implementation is made quantumly using the BB84 protocol. Moreover, private key protocols, once
a secure key is established between Alice and Bob, will not become insecure with the advent of a quantum
computer. This is the main reason that led to the boom of research and development in the field of quantum cryptography
in the last two decades \cite{gis02,sca09,wee12}, culminating with commercially viable quantum cryptography solutions \cite{idq,mqt,ql}.  

The most important concept behind all standard quantum key distribution schemes 
is the use of non-orthogonal quantum states to encode the classical bits $0$ and $1$ \cite{ben92}.
Those bits are randomly generated by Alice and randomly encoded into non-orthogonal quantum states (qubits) 
that are sent to Bob. By properly preparing and measuring those qubits, Alice and Bob are able to share a secret random string of $0$'s
and $1$'s, i.e., the secret key needed to the implementation of private key cryptography 
protocols.\footnote{Alice and Bob must also share an authenticated classical channel, which can be 
totally insecure, to realize the key distribution.}
And since non-orthogonal quantum states cannot be cloned \cite{woo82,die82}, the transmission of
a random sequence of bits using this strategy is secure according to the laws of quantum mechanics.

What happens if the qubits encoding the bits of the secret key 
are not physically sent but rather teleported from Alice to Bob \cite{ben93}? On the one hand
we can think of this process as a substitution to the direct sending of the qubits from Alice to Bob 
through a physical channel. In this case, the teleportation of qubits is an alternative way to send to Bob
the quantum information contained in those qubits, playing no direct role in the generation of the secret key.
Also, no qualitative changes are made to the 
quantum key distribution protocols when we use the quantum teleportation protocol either as a substitute to  
the direct transmission of the qubits or to increase the 
physical distance between Alice and Bob in which they can still establish a secret key \cite{bri98}.

On the other hand, we can use the quantum teleportation protocol as the main ingredient
to the construction of a quantum key distribution protocol, such that the teleportation protocol
plays an \textit{active} role in the generation of the key. This
active role of the teleportation protocol in the generation of the key is the main feature of the quantum key distribution
protocol originally presented in Ref. \cite{rig10}, from now on called the GR10 protocol. In addition to that, the GR10 protocol has two
other interesting features. First, it works securely even when Alice and Bob use 
partially entangled states to implement the quantum teleportation protocol 
\cite{guo00,agr02,rig06,rig06b,gor07,rig09,for13,for15,for16}.
Second, the classical bits of the secret key can be encoded into two \textit{orthogonal} states that are 
subsequently teleported from Alice to Bob. This last feature is in contrast to standard 
quantum key distribution schemes, where the use of non-orthogonal quantum states to encode the classical
bits of the key is mandatory.\footnote{We can also use maximally entangled states to create a secret key shared 
by Alice and Bob without using teleportation and without sending qubits encoding classical bits from Alice to Bob \cite{eke91,ben92b}.} See Refs. \cite{vai95,koa97,shu16} for other quantum key distribution schemes where only orthogonal states are used in its execution.

Although so far no security loophole was found in the GR10 protocol, its complete 
security analysis is lacking. One of the main goals of this work is to fill this gap, proving that
the GR10 protocol as originally conceived is secure against all types of individual and collective
attacks. This is \textcolor{black}{a new contribution of this work whose details are} 
shown in Sec. \ref{gr10}, where we also review how the GR10 protocol works and 
give the main qualitative reasons of why it is indeed secure. \textcolor{black}{Another
new contribution in this work is a discussion on} how the GR10 protocol
can be modified in order to achieve asymptotic unconditional security, i.e.,
security against all types of attacks allowed by the laws of Physics.
But before tackling the security analysis of the GR10 protocol, we revisit in greater detail the security analysis of the
BB84 protocol in Sec. \ref{bb84}, presenting the main tools needed to handle the security analysis of the GR10 protocol. 
We also show that by using the non-uniqueness of the Schmidt decomposition, we can
write the purification of the BB84 protocol in such a way that it is 
possible to prove its security for quantum bit error rates that are higher than those predicted by its standard 
security analysis. \textcolor{black}{This last result, presented in Sec.~{\ref{newBB84}}, is 
new and it adds in our understanding of the secure functioning of the BB84 protocol.}  
\textcolor{black}{The last new result of this work is given} in Sec. \ref{gr10-new}, where
we modify the original GR10 protocol, transforming it into a deterministic
protocol at the cost of a lower secret-key rate.
And finally, in Sec. \ref{conclusion}, we
give our concluding remarks.

\section{Revisiting the security analysis of the BB84 protocol}
\label{bb84}

\subsection{The BB84 protocol}

Before we revisit the security analysis of the BB84 protocol \cite{sho00,lo01}, we give a short description
of how it basically works. It goes as follows:

\begin{itemize}
\item[(i)] In the first place, Alice and Bob agreed beforehand that the encoding of
the classical bits sent from Alice to Bob is randomly done using two 
non-orthogonal quantum states for each bit. For instance,
the bit $0$ is either encoded into the qubit $|0\rangle$ or $|+\rangle = (|0\rangle + |1\rangle)/\sqrt{2}$
and the bit $1$ into $|1\rangle$ or $|-\rangle=(|0\rangle - |1\rangle)/\sqrt{2}$. 
\item[(ii)] After such agreement, Alice randomly prepares her qubit in one of the four states described above and
send it to Bob. After receiving the qubit, Bob measures it randomly using either the z-basis, spanned by the
states $\{ |0\rangle,|1\rangle \}$, or the x-basis, spanned by $\{ |+\rangle,|-\rangle \}$. 
Whenever Bob measurement results are $|0\rangle$ or $|+\rangle$, he assumes Alice sent the bit 0, while
whenever his measurement results are $|1\rangle$ or $|-\rangle$, he assigns the bit value 1.
Step (ii) is repeated $N$ times.
\item[(iii)] After Bob has finished all his measurements on the $N$ qubits, Alice and Bob start a conversation over 
an authenticated  public classical
channel that can be fully insecure. In this step Alice reveals the basis used to prepare the qubits encoding
each bit sent to Bob and he reveals the basis he used to implement his measurements. They discard the cases where different 
bases were used and keep the instances in which they used the same basis. Roughly $N/2$ cases are discarded and $N/2$
are kept. Those $N/2$ remaining shared bits are usually called the raw key and this process of selecting the useful part
of the shared bits from what is not useful is called sifting. Note that the sifting step is crucial. According to
the laws of quantum mechanics, and if there is no noise or eavesdropping, 
Alice and Bob will agree $100\%$ of the time on the value of a given bit whenever Bob measures
his qubit in the same basis used by Alice to prepare it. Had they insisted on keeping 
the cases where different bases were employed, even in principle they could not always agree on the values
of the bits, making a mistake $50\%$ of the time.
\item[(iv)] Alice and Bob then reveal a part of the bits of their raw key, a random sample of
the $N/2$ remaining cases. This sample of bits is used to check for the presence of an
eavesdropper and it usually consists of $N/4$ bits, half the size of the raw key \cite{ben84}. 
If for every pair of bits in the disclosed sample Alice's and Bob's bits are the same, 
Alice and Bob assume no eavesdropping happened and the remaining $N/4$ bits are used as a 
secret key. If, as is always the case in practice, some of the bits disclosed by Alice and Bob
are not the same, they use the disclosed information to estimate the error rate of the 
quantum communication channel. This error rate is always assumed to be caused by Eve's action, whether
or not she really acted on the channel. \textit{Below a certain threshold for the error rate},
Alice and Bob act classically on the remaining undisclosed bits implementing error correction and
privacy amplification protocols. At the end, they share a reduced sequence of bits that can be considered
secure and identical.
\end{itemize}

\subsection{Unconditional security}

One of the main problems of both classical and quantum cryptography is to provide a rigorous 
way to compute the error rate threshold below which a key distribution scheme is secure
\cite{gis02,sca09,wee12}. Specifically, we would like to have an upper bound on the error rate
below which \textit{unconditional} security is achieved, i.e., we want to know the error rate below which 
the key distribution protocol is secure against 
any possible attack that Eve might implement using the known laws of Physics. 

The most general and powerful security attack on a quantum key distribution scheme is usually 
called a coherent attack. Using this terminology, the ultimate goal of quantum cryptography security analysis is thus 
to obtain the tightest upper bound on the error rate below which a given quantum key distribution protocol is
secure against all coherent attacks. 

To understand what a coherent attack is, it is important
to discuss two other types of attacks \cite{gis02,sca09,wee12}. 
The first one is called individual or incoherent attack, the least severe class of attacks.
Individual attacks are such that Eve attacks each one of the $N$ qubits traveling from Alice to Bob independently
of the other qubits and always using the same strategy of attack.\footnote{If Eve uses different strategies, Alice and Bob can detect
Eve by randomly choosing different samples of the raw key to check for security. This will lead to different error rates for different samples
if Eve uses different attacks. Thus, different error rates mean that Eve tampered with the key distribution 
scheme. In this case Alice and Bob discard all the data and restart the key distribution protocol all over.} 
It is also assumed that Eve implements all her measurements on the probes (ancillas) she used to interact
with the qubits sent by Alice before
Alice and Bob start the classical post-processing of the data (error correction and privacy amplification).  

The second type of attack, called collective attack, extends the possibilities of Eve in the sense
that now she can do everything allowed by individual attacks and, in addition, she can decide to 
postpone the measurements on
her ancillas until the moment she considers more convenient, storing the ancillas before any measurement 
in a quantum memory, for example.  Also, 
her measurements can be joint measurements (collective measurements), 
in which she measures more than one ancilla simultaneously. 

The coherent attack, the most powerful attack, is such that Eve can do everything allowed by individual and 
collective attacks, as well as implement any type of interaction 
(unitary operation) involving any number of her ancillas with any number of the qubits sent by Alice
to Bob. Eve is also allowed to implement whatever type of measurements she likes at whatever time she
believes is more convenient to her. In summary, Eve can do with the qubits
sent from Alice to Bob whatever 
is allowed by the laws of quantum mechanics.

Here and in the rest of this work we restrict ourselves to \textit{one-way post-processing} of the raw key. In the BB84 protocol,
this is implemented in the step (iv) described above. One-way post-processing 
is optimally divided into an error correction (information reconciliation) first step and then a 
privacy amplification step \cite{sca09}. These are classical processing protocols applied to the raw key and
they are called one-way whenever only one party sends classical information during the implementation of
those tasks. The other party acts according to previously established rules but never sends
any feedback to the communicating party. Furthermore, if the communicating party is the one that sends the
quantum states, we have \textit{direct reconciliation} one-way post-processing. Using our terminology, Alice is
the party who communicates classically with Bob during the post-processing stage
and he is the one acting on his data without giving any feedback to Alice. If the size (number of bits) of the raw key
is $R$, at the end of the classical post-processing step only a fraction $r$ of the original raw key will be
secure and perfectly correlated. In other words, the shared secret key's size is $K=rR$. 

An important tool in the analysis of the security of a quantum key distribution scheme under collective attacks,
one-way classical post-processing, and direct reconciliation is the Devetak-Winter bound \cite{dev05},
\begin{equation}
r = I(A:B) - \max_{\text{Eve}}{\chi(A:E)}.
\label{dwb}
\end{equation}
Here $r$ is the secret-key fraction as defined above and understood in the asymptotic case, namely,
we are dealing with infinitely long raw keys ($N\rightarrow \infty$) \cite{sca09}. The first term on the 
right hand side of Eq.~(\ref{dwb}),
$I(A:B)$, is the mutual information between the classical data with Alice and Bob, i.e.,
the correlation between the string of bits with Alice and the string of bits with Bob after the sifting step. 
In an ideal scenario, with no noise and no Eve, $r=I(A:B)=1$. This means that every single bit of Alice's bit string
is identical with the corresponding one of Bob's bit string, i.e., the raw key is the secret shared key ($K=R$). 
The second term on the right hand side, $\chi(A:E)$, is the Holevo quantity \cite{hol73}.   
It can be seen as a quantum generalization of the mutual information, quantifying 
Eve's information about the raw key. Maximizing it over all possible collective attacks, 
that is what the notation $\max_{\text{Eve}}$ is telling us to do, we can get the secret-key rate 
of a quantum key distribution protocol by subtracting it from the mutual information of Alice and Bob. 
If $r>0$, the protocol is secure, and if $r\leq 0$, it is insecure \cite{dev05}. We will come back to a more
detailed discussion about $I(A:B)$ and $\chi(A:E)$ later.

The subsequent understanding that a prepare-and-measure quantum key distribution protocol has an
equivalent entanglement-based representation \cite{gis02,sca09,wee12,ben92b,hug93} led to practical ways
to compute the secret-key fraction given by Eq.~(\ref{dwb}). Working directly with the entanglement-based 
version of the key distribution protocol, we can write the quantum state shared by Alice, Bob, and Eve after 
the interference of Eve as a pure state (purification of the state with Alice and Bob). This is the best
scenario for Eve \cite{dev05} and using the purification we can estimate a lower bound for the 
secret-key rate $r$ that depends on the error rates of the protocol \cite{sca09,dev05}. Note that
the mapping of a prepare-and-measure protocol to its entanglement-based representation does not imply
that the latter is equivalently easy to implement in practice as the former. This mapping just tells
us that the security proof obtained for the entanglement-based protocol is as good as if we had
worked directly with the prepare-and-measure protocol \cite{gis02,sca09,wee12}.

The next breakthrough in the security analysis of quantum key distribution protocols,
generalizing the ideas given in Refs. \cite{kra05,kra05b}, was the proof that unconditional security
analysis can be carried out in the asymptotic regime by studying the security of a given protocol 
at the collective attack level \cite{ren07,ren09}. With the help of the quantum de Finetti theorem, 
Renner \cite{ren07}, for discrete variable protocols, and Renner and Cirac \cite{ren09}, for 
continuous variable ones, showed in what sense collective and coherent attack security analysis are equivalent.

For a permutation invariant 
quantum key distribution protocol,
the quantum de Finetti theorem guarantees that in the asymptotic limit Eve's knowledge of the raw key, 
when she implements a coherent attack, is not crucially different from her knowledge 
of it if she had implemented a collective attack. In other words, security under 
collective attacks implies security under coherent attacks for 
any protocol whose entanglement-based representation 
is invariant under any permutation between the quantum states describing the 
$N$ pairs of particles shared by Alice and Bob. And if this permutation invariance is not 
already present due to the symmetry of the protocol, it 
can be enforced at a later stage by a suitable randomization of the classical data 
shared by Alice and Bob \cite{sca09,wee12,ren07,ren09}. 

\subsection{Mutual information and the Holevo quantity}

Before we continue, it is important at this stage to give 
precise definitions of the mutual information \cite{gis02,sca09,wee12,nie00} and of the Holevo quantity \cite{gis02,sca09,wee12,hol73,nie00}.
These two quantities are the main ingredients we need to compute the secret-key fraction, Eq.~(\ref{dwb}), and 
thus assess the security of any quantum key distribution protocol.

\subsubsection{Mutual information}

Let $A$ denote a random variable, $a$ its possible outcomes, and 
$p_A(a)$ the probability of $A$ having the value $a$. The Shannon entropy $H(A)$
associated with the random variable $A$ is \cite{nie00}
\begin{equation}
H(A) = -\sum_{a} p_A(a)\log [p_A(a)],
\label{shannon}
\end{equation}
where the base of the logarithm is $2$ and $0\log 0 = 0$. 
The Shannon entropy quantifies the information content
of the random variable $A$. For a completely random variable whose sample space is composed of $n$ symbols,
we have that $p_A(a)=1/n$, any $a$. This leads to $H(A)$ attaining its maximum value possible: $\log n$. For the trivial 
case where only one $p_A(a)=1$ and all the others are zero, we get $H(A)=0$, the least amount of 
information possible. For any other random distribution of $n$ symbols, $0\leq H(A)\leq \log n$.

If we now similarly introduce another random variable $B$,  the probability of
finding $B$ with the value $b$ given that $A$ has the value $a$ is 
$p_{B|A}(b|a)$. This is usually called the conditional probability. With the aid of
the conditional probability we can define the conditional entropy as
\begin{equation}
H(B|A) = \sum_{a} p_A(a) H(B|A=a),
\label{conditional}
\end{equation}
where
\begin{equation}
H(B|A=a) = -\sum_{b}p_{B|A}(b|a)\log[p_{B|A}(b|a)].
\end{equation}
The conditional entropy quantifies the average information contained on 
the several conditional ``random variables $B|A$'' that can be built for each outcome 
$a$ of the random variable $A$. In a certain sense, it measures the average information needed
to completely characterize the random variable $B$ given our knowledge of $A$. If $A$ and $B$
are perfectly correlated, i.e, if the knowledge of $A$ completely determines $B$, we have that $H(B|A)=0$.
On the other hand, for completely independent variables $A$ and $B$, $H(B|A) = H(B)$.

Using the Shannon and the conditional entropies, the mutual information between the two random variables $A$
and $B$ reads,
\begin{equation}
I(A:B) = H(B) - H(B|A). 
\label{iab}
\end{equation}
The mutual information quantifies the correlation between the random variables $A$ and $B$ and it is related
to the statistical mutual dependence between these two random variables. For completely independent random variables
$A$ and $B$, the mutual information acquires its minimal value possible, $I(A:B)=0$, since $H(B|A) = H(B)$. For perfectly correlated
variables $H(B|A) = 0$ and we thus get $I(A:B)=H(B)$.

\subsubsection{The Holevo quantity}

Let us assume that for each one of the possible outcomes $a$ of a random variable $A$ we associate 
the quantum state $\rho_{E|a}$. We can think of $\rho_{E|a}$ as the quantum state with Eve conditioned on Alice's
obtaining the state $|a\rangle$ after she implements, for instance, a projective measurement on her system. 
The probability $p_A(a)$ of getting the value $a$ from $A$ is such that it is equal  
to Alice's chance of getting after her measurement 
the state $|a\rangle$. In this scenario, the density matrix describing Eve's system is given by 
an ensemble built on the states $\rho_{E|a}$ with probability weight $p_A(a)$,  
\begin{equation}
\rho_{E} = \sum_{a} p_A(a) \rho_{E|a}.
\end{equation}

The state $\rho_{E}$ can also be seen as Eve's description of her system if she knows nothing about the measurement
results obtained by Alice at a given run of a quantum key distribution protocol. This is the case in the BB84 and GR10 protocols,
where the key is built out of the instances in which no information is revealed about Alice's measurement results. Only the 
measurement basis is publicly revealed, which means that Eve has no information about
which state $|a\rangle$ Alice measured.

The state $\rho_E$ is obtained by tracing out Alice's system from the state below, 
which describes Alice and Eve's joint state prior to
any measurement,
\begin{equation}
\rho_{E} = \text{Tr}_A(\rho_{AE}).
\end{equation}

Using $\rho_{AE}$ and the measurement postulate of quantum mechanics, if Alice projects her state
onto $|a \rangle\langle a |$ we have that
\begin{equation}
p_A(a) = \text{Tr}[\rho_{AE}(| a \rangle\langle a |\otimes \mathbb{1}_E)],
\label{pA}
\end{equation}
where we now take the total trace and $\mathbb{1}_E$ is the identity operator acting on Eve's Hilbert space.
The measurement postulate also implies that 
\begin{equation}
\rho_{E|a} = (\langle a |\otimes \mathbb{1}_E)\rho_{AE} (|a \rangle \otimes \mathbb{1}_E)/p_A(a).
\label{rhoE|A}
\end{equation}

The state $\rho_{AE}$ is, in its turn, obtained from the partial trace with respect to Bob of $\rho_{ABE}$, the total
state describing Alice, Bob, and Eve,
\begin{equation}
\rho_{AE} = \text{Tr}_B(\rho_{ABE}).
\label{rhoAE}
\end{equation}

Using the previous notation, the Holevo quantity is given by
\begin{equation}
\chi(A:E) = S(\rho_E) - \sum_{a} p_A(a) S(\rho_{E|a}),
\label{holevoQ}
\end{equation}
where $S(\rho)$ is the von Neumann entropy of the quantum state $\rho$ \cite{nie00},
\begin{equation}
S(\rho) = - \text{Tr}(\rho \log \rho).
\label{vonN}
\end{equation}

We can appreciate the meaning of the Holevo quantity $\chi(A:E)$ 
noting that it bounds the communication capacity of a quantum channel, where 
we encoded the classical symbols $a$, whose probability of occurrence is $p_A(a)$, into the quantum states
$\rho_{E|a}$ \cite{sca09,dev05,hol73}. Because of this property, the 
Holevo quantity is usually considered a quantum extension of the 
mutual information.

\subsection{Entanglement-based representation of the BB84 protocol and its standard security analysis}
\label{ebrBB84}

The entanglement-based representation of the BB84 protocol is \cite{sca09}
\begin{equation}
|\Phi_1\rangle = |\Phi^+\rangle =\frac{1}{\sqrt{2}}(|00\rangle + |11\rangle).
\label{b1}
\end{equation}
This is one of the four Bell states and it reproduces exactly the statistics Alice and Bob 
obtain in the original prepare and measure scheme \cite{ben84}. Indeed, if Alice and Bob measure their qubits in the z-basis,
they will obtain with equal chances either the qubit $|0\rangle$ or the qubit $|1\rangle$. Also, noting that 
$|\Phi^+\rangle=(|++\rangle + |--\rangle)/\sqrt{2}$, a similar perfect correlation follows if they both measure
their qubits in the x-basis. 

The Bell state $|\Phi^+\rangle$ together with the other three,
\begin{eqnarray}
|\Phi_2\rangle = |\Phi^-\rangle& =& \frac{1}{\sqrt{2}}(|00\rangle - |11\rangle),
\label{b2} \\
|\Phi_3\rangle = |\Psi^+\rangle& =& \frac{1}{\sqrt{2}}(|01\rangle + |10\rangle),
\label{b3} \\
|\Phi_4\rangle = |\Psi^-\rangle& =& \frac{1}{\sqrt{2}}(|01\rangle - |10\rangle),
\label{b4} 
\end{eqnarray}
form a complete orthonormal basis that can be used to expand any pure two-qubit state.

In the ideal scenario, after each run of the entanglement-based version of the BB84 protocol, 
Alice and Bob share the state $|\Phi^+\rangle$. 
After $N$ runs they share the state $|\Phi^+\rangle^{\otimes N}$. However, in  the presence of Eve and at 
the collective attack level, Eve attacks each one of the qubits traveling to Bob individually and using the same strategy.\footnote{Although
we will be dealing with collective attacks, the following analysis remains valid for 
the more general class of coherent attacks due to
the results of Refs. \cite{kra05,kra05b,ren07,ren09}.}
This means that the state shared by Alice, Bob, and Eve after $N$ runs of the protocol can be written as 
\begin{equation}
|\Psi \rangle_{ABE}^{\otimes N} = |\Psi \rangle_{ABE} \otimes \cdots \otimes |\Psi \rangle_{ABE},
\end{equation}
where 
\begin{equation}
|\Psi \rangle_{ABE} = \sum_{j=1}^{4}\sqrt{\lambda_j}|\Phi_j\rangle_{AB}|\epsilon_j\rangle_E.
\label{psiABE}
\end{equation}
Equation (\ref{psiABE}) is a purification of the state describing Alice, Bob, and Eve and we can always
use it to describe the global state of our parties due to the Schmidt decomposition theorem \cite{nie00}.
For a large enough Hilbert space describing Eve's states, we can always write $|\Psi \rangle_{ABE}$ as given 
above, with all $\lambda_j$ non negative, $\sum_j\lambda_j = 1$, and $|\epsilon_j\rangle$ forming an orthonormal basis. 

Since after $N$ runs of the protocol the global state shared by all parties is $|\Psi \rangle_{ABE}^{\otimes N}$,
we have an ensemble of $N$ replicas of the state $|\Psi \rangle_{ABE}$. Thus, all correlation between
$AB$ (Alice and Bob) and $E$ (Eve) can be analyzed working directly with $|\Psi \rangle_{ABE}$. Also, 
note that without Eve or noise we have $\lambda_1=1$ and $\lambda_j=0$, $j=2,3,4$. When Eve is present, however,
we have in general that $\lambda_j\neq 0$, for all $j$. 

Using Eq.~(\ref{psiABE}) we are able to determine in which way the presence of Eve affects the BB84 protocol or, equivalently, 
we can study how the presence of noise changes the operation of the BB84 protocol from its expected noiseless case. 
When Eve is present and Alice and Bob 
use the same basis to prepare and measure their qubits, 
the probability of agreement on the value of the bit sent from Alice is no longer one.

Considering the instances in which Alice and Bob employed the z-basis, the probability of agreement on the
value of the bit sent from Alice to Bob is $1-\varepsilon_z =  p_A(0)p_{B|A}(0|0) + p_A(1)p_{B|A}(1|1)$, 
while the probability of Bob making 
a mistake is $\varepsilon_z = p_A(0)p_{B|A}(1|0) + p_A(1)p_{B|A}(0|1)$.
However, a standard result from the theory of probabilities states that $p_{AB}(a,b)=p_A(a)p_{B|A}(b|a)$,
where $p_{AB}(a,b)$ is the joint probability of Alice and Bob seeing simultaneously the values $a$ and $b$,
respectively. This leads to
\begin{equation}
\varepsilon_z = p_{AB}(0,1) + p_{AB}(1,0)
\label{errorZ}
\end{equation}
and to a similar expression for $1-\varepsilon_z$.

Applying the measurement postulate of quantum mechanics we get  
\begin{equation}
p_{AB}(a,b) = \text{Tr}(P_{ab}\rho_{ABE}),
\label{ProbAB}
\end{equation}
where
\begin{equation}
\rho_{ABE} = (|\Psi \rangle_{ABE}) (_{ABE} \langle \Psi |),
\label{rhoABE}
\end{equation}
with $|\Psi \rangle_{ABE}$ given by Eq.~(\ref{psiABE}),
and
\begin{equation}
P_{ab}=|a\rangle_A \,_A\langle a| \otimes |b\rangle_B \,_B\langle b| \otimes \mathbb{1}_E.
\label{PAB}
\end{equation}

Inserting Eqs.~(\ref{rhoABE}) and (\ref{PAB}) into (\ref{ProbAB})  we obtain
\begin{equation}
p_{AB}(0,1) = p_{AB}(1,0) = (\lambda_3 + \lambda_4)/2 
\label{pAB}
\end{equation}
and consequently 
\begin{equation}
\varepsilon_z = \lambda_3 + \lambda_4.
\label{errorZ2}
\end{equation}

If we now consider the cases 
where Alice and Bob employed the x-basis to prepare and measure their qubits, 
a similar calculation leads to 
the following probability of Bob making 
a mistake on the value of the bit sent by Alice, 
\begin{equation}
\varepsilon_x = \lambda_2 + \lambda_4.
\label{errorX}
\end{equation}

We are now ready to compute the two quantities needed to calculate 
the secret-key fraction of the BB84 protocol (see Eq.~(\ref{dwb})).  
We begin with the mutual information and then we move on to the 
Holevo quantity. 

\subsubsection{The mutual information of the BB84 protocol}
\label{mutualBB84}

In order to compute the mutual information between Alice and Bob, we consider the cases in which they 
have employed the z-basis to prepare and measure their qubits. The cases in which
they both use the x-basis will only be employed to estimate the error $\varepsilon_x$ needed to 
the computation of the Holevo quantity \cite{sca09}.  

Due to the symmetry of the state $\rho_{AB}=\text{Tr}_E(\rho_{ABE})$ under 
the permutation of the qubit with Alice with the one with Bob, we have
that $p_A(a)=p_B(a)$. Thus, inserting Eqs.~(\ref{rhoABE}) and (\ref{rhoAE}) into (\ref{pA})
we get
\begin{equation}
p_A(0)=p_A(1) = p_B(0)=p_B(1) = 1/2.
\label{pApB}
\end{equation}

Using Eq.~(\ref{pApB}),	 it is not difficult to see that Eq.~(\ref{shannon}) becomes
\begin{equation}
H(A) = H(B) = 1.
\label{ha}
\end{equation}

Now, using that $p_{AB}(a,b)=p_A(a)p_{B|A}(b|a)$ and that $\sum_b p_{AB}(a,b)=p_A(a)$, it is
possible to compute Eq.~(\ref{conditional}) with the aid of Eqs.~(\ref{pAB}) and (\ref{pApB}). 
This leads to 
\begin{equation}
H(B|A) = h(\varepsilon_z),
\label{hba}
\end{equation}
with
\begin{equation}
h(x) = -x \log x -(1-x)\log (1-x)
\end{equation}
being the binary entropy, and where we have used that  
$p_{AB}(0,0) = p_{AB}(1,1) = (\lambda_1 + \lambda_2)/2$ to write
Eq.~(\ref{hba}) as shown above.

Combining Eqs.~(\ref{ha}) and (\ref{hba}) we finally get the mutual information between
Alice and Bob (see Eq.~(\ref{iab})),
\begin{equation}
I(A:B) = 1 - h(\varepsilon_z).
\label{iab2}
\end{equation}

\subsubsection{The Holevo quantity of the BB84 protocol}

Looking at Eq.~(\ref{holevoQ}), we see that to compute the Holevo 
quantity we need  $\rho_E=\text{Tr}_{AB}(\rho_{ABE})$, Eve's quantum state
after she interacts her probes with the qubit sent by Alice, and $\rho_{E|a}$, Eve's description of her 
physical system if she knows that Alice projected her qubit onto the state $|a\rangle$.

Tracing out Alice and Bob from $\rho_{ABE}$, Eq.~(\ref{rhoABE}), we get
\begin{equation}
\rho_E = \sum_{j=1}^4\lambda_j |\epsilon_j\rangle_{\!E\,E} \langle \epsilon_j |,
\end{equation}
which, after inserted into Eq.~(\ref{vonN}), gives
\begin{equation}
S(\rho_E) = - \sum_{j=1}^4 \lambda_j \log \lambda_j.
\label{SrhoE}
\end{equation}

Furthermore, a direct calculation using Eqs.~(\ref{rhoE|A}), (\ref{rhoAE}), (\ref{rhoABE}), 
and (\ref{pApB}) leads to 
\begin{eqnarray}
\rho_{E|0} &=& \sum_{j=1}^{4}\lambda_j \ket{\epsilon_j}_{\!E\,E\!}\bra{\epsilon_j} 
+ \sqrt{\lambda_1 \lambda_2}\left(\ket{\epsilon_1}_{\!E\,E\!}\bra{\epsilon_2} 
+ h.c.\right) \nonumber \\
&& + \sqrt{\lambda_3 \lambda_4}\left(\ket{\epsilon_3}_{\!E\,E\!}\bra{\epsilon_4} 
+ h.c.\right),  \\
\label{rhoE|0}
\rho_{E|1} &=& \sum_{j=1}^{4}\lambda_j \ket{\epsilon_j}_{\!E\,E\!}\bra{\epsilon_j} 
- \sqrt{\lambda_1 \lambda_2}\left(\ket{\epsilon_1}_{\!E\,E\!}\bra{\epsilon_2} 
+ h.c.\right) \nonumber \\
&& - \sqrt{\lambda_3 \lambda_4}\left(\ket{\epsilon_3}_{\!E\,E\!}\bra{\epsilon_4} 
+ h.c.\right), 
\label{rhoE|1}
\end{eqnarray}
where $h.c.$ denotes the Hermitian conjugate of the operator appearing before it.
Noting that the eigenvalues of both $\rho_{E|0}$ and $\rho_{E|1}$ are $0,0,\lambda_1+\lambda_2$,
and $\lambda_3+\lambda_4$, the von Neumann entropy, Eq.~(\ref{vonN}), for those states becomes
\begin{equation}
S(\rho_{E|0}) = S(\rho_{E|1}) = h(\varepsilon_z),
\label{SrhoE0}
\end{equation}
where we have used Eq.~(\ref{errorZ2}) to eliminate the $\lambda$'s in favor of the 
error rate $\varepsilon_z$.

Inserting Eqs.~(\ref{SrhoE}) and (\ref{SrhoE0}) into (\ref{holevoQ}), and using that 
$p_A(a)=1/2$, the Holevo quantity reads
\begin{equation}
\chi(A:E) = - \sum_{j=1}^4 \lambda_j \log \lambda_j - h(\varepsilon_z).
\label{chiAE}
\end{equation}

\subsubsection{The secret-key fraction of the BB84 protocol}

If we insert Eqs.~(\ref{iab2}) and (\ref{chiAE}) into (\ref{dwb}), we get the following expression for 
the secret-key fraction of the BB84 protocol,
\begin{eqnarray}
r &=&  1 - h(\varepsilon_z) - \max_{\text{Eve}}\left\{- \sum_{j=1}^4 \lambda_j \log \lambda_j - h(\varepsilon_z)\right\} \nonumber \\
&=& 1 - h(\varepsilon_z) + h(\varepsilon_z) - \max_{\text{Eve}}\left\{- \sum_{j=1}^4 \lambda_j \log \lambda_j \right\} \nonumber \\
&=& 1 + \min_{\text{Eve}}\left\{\sum_{j=1}^4 \lambda_j \log \lambda_j \right\}.
\label{rBB84}
\end{eqnarray}
Note that to arrive at the last equality we used that $$-\max_{x}[f(x)]=\min_{x}[-f(x)],$$ i.e.,
the negative of the maximum value of a function $f$ over its domain is equal to the minimum of $-f$ over
the same domain.

For the original BB84 protocol, we have shown that the $\lambda$'s satisfy the following three equations,
\begin{eqnarray}
\lambda_3 + \lambda_4 &=& \varepsilon_z, \label{erroZ}\\
\lambda_2 + \lambda_4 &=& \varepsilon_x, \label{erroX}\\
\sum_{i=1}^4 \lambda_i &=& 1. \label{normalization} 
\end{eqnarray}
Since we have four $\lambda$'s, we cannot uniquely express them in terms of the 
experimentally determined error rates $\varepsilon_z$ and $\varepsilon_x$. This means 
that we will need to minimize the expression inside the curly brackets in Eq.~(\ref{rBB84}) 
in order to get a lower bound of the secret-key fraction as a function of
those error rates. This is most easily done following the prescription given in Ref. \cite{sca09}.

Using Eq.~(\ref{erroZ}) we see that $\lambda_4=\varepsilon_z - \lambda_3$ and thus   
that $0\leq \lambda_3\leq \varepsilon_z$ since no $\lambda_j$ can be negative.
This means that we can write 
\begin{eqnarray}
\lambda_3&=&v\varepsilon_z, \label{L1} \\
\lambda_4&=&(1-v)\varepsilon_z, \label{L2}
\end{eqnarray}
where $v\in[0,1]$.

Now, Eq.~(\ref{normalization})
can be written as $\lambda_2 =1-\lambda_1-\lambda_3-\lambda_4 = (1 - \varepsilon_z)-\lambda_1$, where we used 
Eq.~(\ref{erroZ}) to arrive at the last equality. Since $\lambda_2$ cannot be negative we have
$0\leq \lambda_1\leq 1-\varepsilon_z$ and, as before, we can write 
\begin{eqnarray}
\lambda_1 &=& u(1-\varepsilon_z), \label{L3} \\ 
\lambda_2 &=& (1-u)(1-\varepsilon_z), \label{L4}
\end{eqnarray}
with $u\in[0,1]$. 

The parameters $u$ and
$v$ are not independent since Eq.~(\ref{erroX}) implies that
%
$
(1-u)(1-\varepsilon_z) + (1-v)\varepsilon_z = \varepsilon_x.
$
%
Solving for $u$ we get
\begin{equation}
u = (1-\varepsilon_x-v\varepsilon_z)/(1-\varepsilon_z),
\label{uv}
\end{equation}
leaving us with only $v$ as a free parameter. Inserting Eq.~(\ref{uv}) 
into (\ref{L1}) and (\ref{L2}) and using Eqs.~(\ref{L1})-(\ref{L3}), we can
write the secret-key fraction (\ref{rBB84}) as 
\begin{equation}
r = 1 + \min_{\text{Eve}}\left\{ \theta\left(1-\varepsilon_x-\varepsilon_z v\right)
+\theta\left(\varepsilon_x-\varepsilon_z (1-v)\right) +\theta(\varepsilon_z-\varepsilon_z v)+\theta(\varepsilon_z v) \right\},
\label{rBB84-2}
\end{equation}
where
\begin{equation}
\theta(x) = x\log x.
\end{equation}

Minimizing $r$ as a function of $v$, i.e., solving  
$dr/dv = 0$ gives
\begin{equation}
v = 1-\varepsilon_x,
\label{vmin}
\end{equation}
which is indeed a minimum for $r$ since a direct calculation leads to $d^2r/dv^2>0$ when
$v=1-\varepsilon_x$.

Using Eqs.~(\ref{uv}) and (\ref{vmin}) we can express the $\lambda$'s giving the lower bound
for the secret-key fraction as
\begin{eqnarray}
\lambda_1    &=&    (1-\varepsilon_x)(1-\varepsilon_z),  \label{L1o}  \\
\lambda_2    &=&    \varepsilon_x(1-\varepsilon_z),	\\
\lambda_3    &=&    \varepsilon_z(1-\varepsilon_x),	\\
\lambda_4    &=&    \varepsilon_z \varepsilon_x. \label{L4o}		
\end{eqnarray}
Inserting Eqs.(\ref{L1o})-(\ref{L4o}) into (\ref{rBB84}) we obtain the lower bound for the 
secret-key fraction of the BB84 protocol in terms of the measurable quantities $\varepsilon_z$ and $\varepsilon_x$,
\begin{equation}
r = 1-h(\varepsilon_x) - h(\varepsilon_z).
\end{equation}
The quantities $\varepsilon_x$ and $\varepsilon_z$ 
are the probability of Bob making a mistake (error rate) about the value of a bit sent by Alice 
when they both use the same basis,
either the x- or  z-basis, to prepare and measure the qubit encoding that bit.

If we assume that $\varepsilon_x=\varepsilon_z=\varepsilon$, the secret-key fraction can be written as
\begin{equation}
r = 1 - 2h(\varepsilon),
\end{equation}
where it is not difficult to see that $r>0$ whenever
\begin{equation}
\varepsilon \lesssim 11\%.
\end{equation}
In other words, for symmetrical error rates below $11\%$ the BB84 protocol can be considered secure \cite{sca09}.

\subsection{Non-uniqueness of the Schmidt decomposition and the security of the BB84 protocol}
\label{newBB84}

Instead of Eq.~(\ref{psiABE}), a perfectly legitimate Schmidt decomposition representing the
state of Alice, Bob, and Eve is
\begin{equation}
|\tilde{\Psi} \rangle_{ABE} = \sum_{j=1}^{4}\sqrt{\tilde{\lambda}_j}|\tilde{\Phi}_j\rangle_{AB}|\tilde{\epsilon}_j\rangle_E,
\label{psiABE2}
\end{equation}
where
\begin{eqnarray}
|\tilde{\Phi}_1\rangle & =& \frac{1}{\sqrt{2}}(|00\rangle + |11\rangle),
\label{b1n} \\
|\tilde{\Phi}_2\rangle & =& \frac{1}{\sqrt{2}}(|00\rangle - |11\rangle),
\label{b2n} \\
|\tilde{\Phi}_3\rangle & =& |01\rangle,
\label{b3n} \\
|\tilde{\Phi}_4\rangle & =& |10\rangle,
\label{b4n}
\end{eqnarray}
span a complete orthonormal basis that can be used to describe any two-qubit pure state. Note that 
$|\tilde{\Phi}_1\rangle=|\Phi_1\rangle$ and $|\tilde{\Phi}_2\rangle=|\Phi_2\rangle$, where
$|\Phi_1\rangle$ and $|\Phi_2\rangle$ are given by Eqs.~(\ref{b1}) and (\ref{b2}). Moreover, when Eve is
absent $\tilde{\lambda}_1=1$, with the other $\tilde{\lambda}$'s being zero, recovering the entanglement-based representation
of the BB84 protocol in the ideal scenario. 

Equation (\ref{psiABE2}) is another purification describing the state of Alice, Bob, and Eve after Eve has 
coupled her ancillas with the qubit sent from Alice to Bob. The same arguments that led us to write Eq.~(\ref{psiABE})
apply here and, as before, $0\leq \tilde{\lambda}_j \leq 1$, $\sum_j\tilde{\lambda}_j = 1$, 
and $|\tilde{\epsilon}_j\rangle$, $j=1,\ldots, 4$, span an orthonormal basis. 

The purifications (\ref{psiABE}) and (\ref{psiABE2}) are connected by the following relation
\begin{equation}
|\tilde{\Psi} \rangle_{ABE} =(U_{AB}\otimes U_E)|\Psi \rangle_{ABE}, 
\end{equation}
where
\begin{eqnarray}
U_{AB} &=& \sum_{j=1}^2\ket{\Phi_j}_{\!\!AB\,AB\!\!}\bra{\Phi_j} 
\!+\! \ket{01}_{\!\!AB\,AB\!\!}\bra{\Phi_3} \!+\! \ket{10}_{\!\!AB\,AB\!\!}\bra{\Phi_4}\!,
\\
U_E &=& \sum_{j=1}^4\ket{\tilde{\epsilon}_j}_{\!\!E\,E\!\!}\bra{\epsilon_j}. 
\end{eqnarray}
It is worth mentioning that the unitary operation connecting the two purifications is local with respect to the partition
$AB$ and $E$. 

Since this unitary operation is local with respect to Eve, we can 
alternatively go from (\ref{psiABE}) to (\ref{psiABE2}) by identifying 
$|\tilde{\epsilon}_j\rangle$ with $|\epsilon_j\rangle$, $\tilde{\lambda}_j$ with $\lambda_j$, 
and by simply applying the following unitary operation,
\begin{equation}
U_{AB}\otimes \mathbb{1}_E,
\end{equation}
where $\mathbb{1}_E$ is the identity operator acting on Eve's Hilbert space. Since
Eve's unitary
operation is given by the identity operator, it is clear that no changes are made on the states describing her system.
Also, whether we think of $|\tilde{\Psi} \rangle_{ABE}$ as given by
$U_{AB}\otimes U_E|\Psi \rangle_{ABE}$ or by $U_{AB}\otimes \mathbb{1}_E|\Psi \rangle_{ABE}$,
we will get the same lower bound for the secret-key fraction since we need to compute the Holevo quantity
maximizing it over all possible strategies that Eve might employ. As such, and in order to simplify notation,
we will write $|\epsilon_j\rangle$ and $\lambda_j$ instead of $|\tilde{\epsilon}_j\rangle$ and $\tilde{\lambda}_j$.

However, the purification (\ref{psiABE2}) is not equivalent to (\ref{psiABE}) in the following sense.  
As we will show below, the way we write the purification (\ref{psiABE2}) forces us to introduce 
another constraint among the $\lambda$'s that is absent from (\ref{psiABE}). This will allow 
us to get a tighter lower bound for the secret-key fraction, which ultimately leads
to an increase of the error rates below which the BB84 protocol continues to operate securely.

Let us now present this extra constraint. 
If we repeat the steps that led to Eq.~(\ref{pApB}) using Eq.~(\ref{psiABE2}) instead of 
(\ref{psiABE}) we get,
\begin{eqnarray}
p_A(0) &=& (\lambda_1+\lambda_2)/2 + \lambda_3, \label{pa0}\\
p_A(1) &=& (\lambda_1+\lambda_2)/2 + \lambda_4. \label{pa1}
\end{eqnarray}
But whether or not Eve interferes, the BB84 protocol is such that we always have
\begin{equation}
p_A(0)=p_A(1)=1/2,
\label{meio}
\end{equation}
since Alice randomly chooses with equal chances whether she sends Bob the bit $0$ or $1$.
Enforcing the condition (\ref{meio}) onto Eqs.~(\ref{pa0}) and (\ref{pa1}), we see that we must
have
\begin{equation}
\lambda_3 = \lambda_4 = \lambda.
\label{constraint}
\end{equation}

Using the constraint (\ref{constraint}) it is not difficult to see that a direct calculation, similar to 
what we did in Sec. \ref{ebrBB84}, leads to
\begin{equation}
p_A(0)=p_A(1) = p_B(0)=p_B(1) = 1/2
\label{pApB2}
\end{equation}
and to
\begin{eqnarray}
p_{AB}(0,0) = p_{AB}(1,1) &=& (\lambda_1+\lambda_2)/2, \\
p_{AB}(0,1) = p_{AB}(1,0) &=& \lambda.
\end{eqnarray}

Similarly, we also obtain for the mutual information between Alice and Bob,
\begin{equation}
I(A:B) = 1 - h(\varepsilon_z),
\label{iabNovo}
\end{equation}
where
\begin{eqnarray}
\varepsilon_z &=& 2\lambda, \label{novoEz}\\
\varepsilon_x &=& \lambda_2 + \lambda. \label{novoEx}
\end{eqnarray}

We now move on to the calculation of the Holevo quantity, Eq.~(\ref{holevoQ}), using the
new purification (\ref{psiABE2}). Using Eq.~(\ref{psiABE2}) and the constraint 
(\ref{constraint}) we get
\begin{displaymath}
\rho_E =  \text{Tr}_{AB}(\rho_{ABE}) = \sum_{j=1}^2\lambda_j |\epsilon_j\rangle_{\!E\,E\!} \langle \epsilon_j |
+ \lambda (|\epsilon_3\rangle_{\!E\,E\!} \langle \epsilon_3 |+|\epsilon_4\rangle_{\!E\,E\!} \langle \epsilon_4 |)
\end{displaymath}
and thus 
\begin{equation}
S(\rho_E) = - \lambda_1\log \lambda_1 - \lambda_2 \log \lambda_2 - 2\lambda\log\lambda.
\label{SrhoE2}
\end{equation}

Now, employing Eq.~(\ref{psiABE2}) instead of (\ref{psiABE}) we get after 
Eqs.~(\ref{rhoE|A}), (\ref{rhoAE}), (\ref{rhoABE}), 
and (\ref{pApB}), 
\begin{eqnarray}
\rho_{E|0} &=& \lambda_1 \ket{\epsilon_1}_{\!E\,E\!}\bra{\epsilon_1} 
+ \lambda_2 \ket{\epsilon_2}_{\!E\,E\!}\bra{\epsilon_2} 
+ 2\lambda\ket{\epsilon_3}_{\!E\,E\!}\bra{\epsilon_3} \nonumber \\
&&+ \sqrt{\lambda_1 \lambda_2}\left(\ket{\epsilon_1}_{\!E\,E\!}\bra{\epsilon_2} 
+ h.c.\right),  \\
\label{rhoE|02}
\rho_{E|1} &=& \lambda_1 \ket{\epsilon_1}_{\!E\,E\!}\bra{\epsilon_1} 
+ \lambda_2 \ket{\epsilon_2}_{\!E\,E\!}\bra{\epsilon_2} 
+ 2\lambda\ket{\epsilon_4}_{\!E\,E\!}\bra{\epsilon_4} \nonumber \\
&& - \sqrt{\lambda_1 \lambda_2}\left(\ket{\epsilon_1}_{\!E\,E\!}\bra{\epsilon_2} 
+ h.c.\right). 
\label{rhoE|12}
\end{eqnarray}
The eigenvalues of both $\rho_{E|0}$ and $\rho_{E|1}$ are $0,0,\lambda_1+\lambda_2$,
and $2\lambda$, leading to the following value for the von Neumann entropy,
\begin{equation}
S(\rho_{E|0}) = S(\rho_{E|1}) = h(\varepsilon_z),
\label{SrhoE02}
\end{equation}
after using Eq.~(\ref{novoEz}).

The Holevo quantity is obtained when we insert Eqs.~(\ref{pApB2}), (\ref{SrhoE2}), and (\ref{SrhoE02}) into (\ref{holevoQ}),
giving
\begin{equation}
\chi(A:E) = - \lambda_1\log \lambda_1 - \lambda_2 \log \lambda_2 - 2\lambda\log\lambda - h(\varepsilon_z).
\label{chiAE2}
\end{equation}
Finally, the secret-key fraction, Eq.~(\ref{dwb}), 
comes from subtracting Eq.~(\ref{chiAE2}) from (\ref{iabNovo}),
\begin{eqnarray}
r &=&  1 +\lambda_1\log \lambda_1 + \lambda_2 \log \lambda_2 + 2\lambda\log\lambda.
\label{rBB84novo}
\end{eqnarray}

It is worth mentioning that we do not need this time 
to minimize $r$ over all strategies that Eve might employ because all quantities are 
already fixed by the experimental data, i.e., all we need to compute $r$ is $\varepsilon_x$ and $\varepsilon_z$.
This comes about due to the extra constraint, Eq.~(\ref{constraint}), which 
allows us to uniquely determine the 
$\lambda$'s as functions of the error rates $\varepsilon_x$ and $\varepsilon_z$. Using Eqs.~(\ref{novoEz}), 
(\ref{novoEx}), and the normalization condition of the $\lambda$'s, namely, $\lambda_1+\lambda_2+2\lambda =1$,
we have
\begin{eqnarray}
\lambda_1    &=&    1 - \varepsilon_x - \varepsilon_z/2,  \label{L1n}  \\
\lambda_2    &=&    \varepsilon_x -\varepsilon_z/2,	\\
\lambda      &=&    \varepsilon_z/2.	\label{Ln}		
\end{eqnarray}

The secret-key fraction is obtained inserting Eqs.~(\ref{L1n})-(\ref{Ln})
into (\ref{rBB84novo}), which can be written after some algebra as follows,
\begin{equation}
r =  \varepsilon_z\log\varepsilon_z + (1-\varepsilon_x-\varepsilon_z/2)\log(2-2\varepsilon_x-\varepsilon_z)
+ (\varepsilon_x-\varepsilon_z/2)\log(2\varepsilon_x-\varepsilon_z).
\label{rBB84novo2}
\end{equation}

Assuming, as before, we deal with a depolarizing channel 
($\varepsilon_x=\varepsilon_z=\varepsilon$) the secret-key fraction becomes
\begin{equation}
r = (3\varepsilon/2)\log\varepsilon + (1-3\varepsilon/2)\log(2-3\varepsilon).
\label{rBB84novo2b}
\end{equation}
Searching for the maximal $\varepsilon$ giving $r>0$ we get
\begin{equation}
\varepsilon \lesssim 12.61\%.
\end{equation}
The above error rate threshold is the same one obtains working with the six-state protocol \cite{sca09,ben84b,bru98,bec99,lo01},
where in addition to the z- and x-basis we have the bits encoded in the y-basis.
This means that both the BB84 and the six-state protocols are secure under the same level of noise.	
It is worth stressing once more that we could only achieve the previous result 
due to the exploration of the non-uniqueness of the purification (Schmidt decomposition) describing the state
of Alice, Bob, and Eve. 

We also point out that we have carried out the security analysis for the BB84 protocol using the Schmidt decomposition 
(\ref{psiABE2}) without imposing the constraint (\ref{constraint}). In this case we had to minimize $r$ over all possible
strategies of Eve because there still remained a free $\lambda$ in the expression for $r$.
At the end of the minimization process we found out that the constraint (\ref{constraint}) naturally emerged as the scenario
that is best to Eve. Since the constraint (\ref{constraint}) is related to the fact that  
$p_A(0)=p_A(1) = 1/2$, it is possible that a non-symmetrical protocol, such that $p_A(0)\neq p_A(1)$,
may lead to an increase in the level of noise in which it still operates securely.

\section{Asymptotic security analysis of the GR10 protocol}
\label{gr10}

To better appreciate all the steps involved in the operation of the GR10 protocol \cite{rig10} and to
get a qualitative feeling of why it is secure, it is crucial to understand 
how the probabilistic teleportation protocol works \cite{guo00,agr02,rig06,rig09,for13}. 
A brief review of the probabilistic teleportation protocol, 
gauged to the purposes of this work, is the subject of Sec. \ref{gr10-a}. 
In Sec. \ref{gr10-b} we present the GR10 protocol using the notation employed in this work.
In Sec. \ref{gr10-c} we start by highlighting the main qualitative  
features of the GR10 protocol that makes it secure against an eavesdropper attack
and then we 
provide its full and rigorous security analysis, similar in spirit to the ones
previously shown for the BB84 protocol.

\subsection{The probabilistic teleportation protocol}
\label{gr10-a}

Alice's goal is to teleport to Bob the state
\begin{equation}
|\phi\rangle_A=\alpha|0\rangle_A + \beta|1\rangle_A
\label{qubitA}
\end{equation}
through the following entangled state shared with him,   
\begin{equation}
|\Phi_1^n\rangle_{AB}=\frac{|00\rangle_{AB}
+n|11\rangle_{AB}}{\sqrt{1+n^2}}.
\label{phi1n}
\end{equation}
Here $\alpha,\beta$ are complex numbers, $|\alpha|^2+|\beta|^2=1$ is the normalization condition,
and $0\leq n\leq 1$. Note that Eq.~(\ref{phi1n}) is not a maximally entangled state whenever
$n<1$. Equation (\ref{phi1n}) together with the generalized Bell states \cite{guo00,agr02,rig06} 
\begin{eqnarray}
|\Phi_2^n\rangle&=&\frac{n|00\rangle -|11\rangle}{\sqrt{1+n^2}}, \\
|\Phi_3^n\rangle&=&\frac{|01\rangle +n|10\rangle}{\sqrt{1+n^2}}, \\
|\Phi_4^n\rangle&=&\frac{n|01\rangle - |10\rangle}{\sqrt{1+n^2}},
\label{basegeneralizada}
\end{eqnarray}
form a complete set of orthonormal vectors spanning a two-qubit Hilbert space.

All the steps of the probabilistic teleportation protocol are similar to the deterministic 
one \cite{ben93}. The two differences are related to the use of a non-maximally entangled 
state connecting Alice and Bob, $|\Phi_1^n\rangle_{AB}$,  and to the basis onto which Alice's projects
her qubits. Instead of projecting her qubits onto maximally entangled Bell states
($n=1$), she projects them onto generalized Bell states $|\Phi_j^m\rangle_{AA}$, where we
assume for the moment that $m\neq n$.

The probability $p_j$ of Alice measuring a particular generalized Bell state $|\Phi_j^m\rangle$ is
\begin{equation}
p_1 = [f_1(\alpha,\beta)]^2,\;\; p_2 = [f_2(\alpha,\beta)]^2,\;\; 
p_3 = [f_2(\beta,\alpha)]^2,\;\; p_4 = [f_1(\beta,\alpha)]^2, 
\end{equation}
where
\begin{equation}
f_1(\alpha,\beta)=\sqrt{\frac{|\alpha|^2+m^2n^2|\beta|^2}
{(1+m^2)(1+n^2)}},\;\; 
f_2(\alpha,\beta)=\sqrt{\frac{m^2 |\alpha|^2+n^2|\beta|^2}
{(1+m^2)(1+n^2)}}.
\end{equation}

After implementing her measurement, Alice tells Bob via a classical communication channel 
the result she obtained. If she measured the state $|\Phi_j^m\rangle$, Bob's state collapses to
$U^\dagger_j|\phi_j\rangle_B$, with $|\phi_j\rangle_B$ given by Eqs.~(\ref{phi1})-(\ref{phi4}). To
finish the teleportation protocol, Bob uses the information received from Alice to apply 
the corresponding unitary operation $U_j$ on his qubit. Here $U_1 = \mathbb{1}$ is the identity matrix
and $U_2=\sigma_{z},
U_3=\sigma_{x}$, and $U_4 = \sigma_{z}\sigma_{x}$ are given by the standard Pauli matrices.   
After applying $U_j$ his qubit is given by one of these four possibilities,
\begin{eqnarray}
|\phi_1\rangle_B=\frac{\alpha|0\rangle_B + mn\beta|1\rangle_B}
{\sqrt{|\alpha|^2+m^2 n^2 |\beta|^2}}, \label{phi1} \\
|\phi_2\rangle_B=\frac{m \alpha|0\rangle_B + n\beta|1\rangle_B}
{\sqrt{m^2 |\alpha|^2+n^2 |\beta|^2}},\label{phi2} \\
|\phi_3\rangle_B=\frac{n\alpha|0\rangle_B + m \beta|1\rangle_B}
{\sqrt{n^2 |\alpha|^2+m^2 |\beta|^2}},\label{phi3} \\
|\phi_4\rangle_B=\frac{m n \alpha|0\rangle_B + \beta|1\rangle_B}
{\sqrt{m^2 n^2 |\alpha|^2+|\beta|^2}}.
\label{phi4}
\end{eqnarray}
In other words, if Alice obtained the state $|\Phi_j^m\rangle$
after projecting her two qubits onto the generalized Bell states,
Bob's qubit at the end of the teleportation protocol 
is correspondingly described by the state $|\phi_j\rangle_B$.
It is worth mentioning that for $m=n=1$ we get back to 
the original teleportation protocol \cite{ben93},  where $p_j=1/4$ and 
$|\phi_j\rangle_B=\alpha|0\rangle_B+\beta|1\rangle_B$ for any $j$.

Looking at Eqs.~(\ref{phi1})-(\ref{phi4}), we realize that Bob
gets a perfect replica of Alice's teleported qubit if she chooses
$m=n$. This condition for the values of $m$ and $n$ is often called \textit{the matching condition} \cite{rig10,agr02,rig06,rig09,for13}
and it is clear that only two out of the four possible measurement results of Alice lead to a perfect teleportation.
This happens whenever she obtains either $|\Phi_2^m\rangle$ or $|\Phi_3^m\rangle$, with $m=n$. In this scenario,
$|\phi_2\rangle_B$ and $|\phi_3\rangle_B$ are given by 
$\alpha|0\rangle_B + \beta|1\rangle_B$ and the teleportation protocol is considered successful. The 
probability of success is  
\begin{equation}
p_{suc}(n)=p_2+p_3=\frac{2n^2}{(1+n^2)^2}.
\label{psuc}
\end{equation}

\subsection{The GR10 protocol}
\label{gr10-b}

Once the probabilistic teleportation protocol is understood, we are ready to show how the GR10 protocol works. 
In what follows we will briefly review all the steps in the execution of the GR10 protocol, similarly to the way 
Ref. \cite{rig10} originally introduced it. 

\begin{itemize}
\item[(i)] Alice and Bob agree beforehand on two points. First, the encoding of the classical bits will be made using only one
orthonormal basis. This basis is chosen such that its base vectors are non-orthogonal to the base vectors used to express the
partially entangled Bell states employed in the probabilistic teleportation protocol. For instance, if the 
entangled resource shared between Alice and Bob is given by 
$|\Phi_1^n\rangle_{AB}=(|00\rangle_{AB}
+n|11\rangle_{AB})/\sqrt{1+n^2}$, they agree that the qubit 
$|+\rangle = (|0\rangle + |1\rangle)/\sqrt{2}$ encodes the bit $0$ and 
$|-\rangle=(|0\rangle - |1\rangle)/\sqrt{2}$ the bit $1$. In other words, they use the x-basis to encode the bits and 
the z-basis to write the Bell states.\footnote{We can always use more than one orthonormal basis to encode 
the classical bits, with non-orthogonal states encoding the same bit. 
If we use two such basis, we can see the present protocol as an additional security layer to the 
BB84 protocol. But the whole point of the GR10 protocol, which will be made clearer when we present its security analysis,
is that it is secure even if we use only one orthogonal basis to encode the classical bits.} 
Second, they also agree on the possible values of $n$. In the original version of the GR10 protocol, only two 
values for $n$ are used, $n_1$
and $n_2$. 
\item[(ii)] Alice randomly prepares a qubit in one of the two states described above, namely,
$|+\rangle$ or $|-\rangle$. Bob, in its turn, randomly generates either the partially entangled two-qubit state
$|\Phi_1^{n_1}\rangle_{AB}$ or $|\Phi_1^{n_2}\rangle_{AB}$, sending one of the qubits to Alice and keeping the
other with him. 
\item[(iii)]
After receiving her share of the partially entangled state, Alice initiates the 
probabilistic teleportation protocol as previously described.
Specifically, she projects the qubit encoding the classical bit 
and the qubit received from Bob
onto the generalized Bell state $|\Phi_j^m\rangle_{AA}$, where 
$m$ is randomly chosen between $n_1$ or $n_2$.  Note that at this stage of the protocol,
\textit{neither Bob tells Alice the value of $n$} he chose to prepare the entangled state
\textit{nor Alice tells Bob the value of $m$} she chose in order to fix the generalized Bell basis
used to project her qubits.  
Alice only tells Bob the 
generalized Bell state $|\Phi_j^m\rangle_{AA}$ she got, $j=1,\ldots,4$, but not the value of $m$. 
\item[(iv)] After receiving the news from Alice 
(the value of $j$ but not of $m$),
Bob implements the corresponding unitary operation to correct his
qubit as described in Sec. \ref{gr10-a}. His qubit, after that, is in one of
the four possible states listed in Eqs.~(\ref{phi1})-(\ref{phi4}), where
$\alpha=|\beta|=1/\sqrt{2}$. He then projects his qubit onto the x-basis. 
If his measurement result is $|+\rangle$, 
he assumes Alice sent the bit 0, and if it yields the state $|-\rangle$,
he assigns the bit value 1. This whole process, steps (i) to (iv), is repeated 
$N$ times.
\item[(v)] After Bob has finished all his measurements on the $N$ qubits, Alice and Bob, similarly to 
what is done in the BB84 protocol, start a conversation over 
an authenticated public classical
channel that can be fully insecure. 
In this public discussion, they reveal to each other the following 
pieces of information. He tells Alice the values of $n$ ($n_1$ or $n_2$) he employed to
prepare each one of the partially entangled states and Alice tells him the values of
$m$ ($n_1$ or $n_2$) used in each one of the $N$ generalized Bell measurements she made. 
They discard the instances in which $m\neq n$, keeping only the cases where the matching condition 
occurred ($m=n$). About $N/2$ cases are discarded and $N/2$
are kept. Of the remaining $N/2$ cases, another sifting process is implemented. 
Due to the particular operation of the probabilistic teleportation protocol, the teleported
qubit reaching Bob is an exact replica of Alice's qubit if $m=n$ and her measurement 
outcome is either
$|\Phi_2^n\rangle$ or $|\Phi_3^n\rangle$. Therefore, they discard the cases in which 
Alice's measurement resulted in the states $|\Phi_1^n\rangle$ and $|\Phi_4^n\rangle$.
The remaining bits are the raw key which, after Eq.~(\ref{psuc}), is given by \cite{rig10} 
\begin{equation}
R=\left(\frac{p_{suc}(n_1)}{2}+\frac{p_{suc}(n_2)}{2}\right)\frac{N}{2}.
\end{equation}
\item[(vi)] Finally, Alice and Bob disclose part of the bits of the raw key, which 
is used to detect the presence of an
eavesdropper. The publicly revealed information is employed to estimate the error rate of the 
quantum communication channel, allowing them to act accordingly to increase the security of the shared 
key via classical reconciliation and privacy amplification protocols. 
\end{itemize}

\subsection{Security analysis of the GR10 protocol}
\label{gr10-c}

The simplest way to qualitatively understand the security of the GR10 protocol is to push as far as possible
the analogy between the GR10 and the BB84 protocols \cite{rig10}. To do this, we first observe that in the GR10
protocol Alice and Bob discard all the instances in which her generalized Bell measurements yield 
$|\Phi_1^m\rangle$ and $|\Phi_4^m\rangle$. Of the remaining two possibilities, 
four different cases emerge. Alice can either teleport the state $\ket{+}$ and obtain $|\Phi_2^m\rangle$ for her
Bell measurement or she can teleport $\ket{+}$ and obtain $|\Phi_3^m\rangle$. Similarly, she can choose to 
teleport to Bob the state $\ket{-}$ and either obtain $|\Phi_2^m\rangle$ or $|\Phi_3^m\rangle$ after
implementing the generalized Bell measurement. The quantum states describing Bob's qubit for each one these four possibilities are respectively 
(see Eqs.~(\ref{phi2}) and (\ref{phi3})),
\begin{eqnarray}
\!\ket{\widetilde{0}}_B \!=\! \frac{m\ket{0}+n\ket{1}}{\sqrt{n^2+m^2}} \!=\! \frac{(m+n)\ket{+}+(m-n)\ket{-}}{\sqrt{2(n^2+m^2)}}, \label{0l}\\
\!\ket{\widetilde{+}}_B \!=\! \frac{n\ket{0}+m\ket{1}}{\sqrt{n^2+m^2}} \!=\! \frac{(m+n)\ket{+}+(n-m)\ket{-}}{\sqrt{2(n^2+m^2)}}, \label{+l}\\
\!\ket{\widetilde{-}}_B \!=\! \frac{m\ket{0}-n\ket{1}}{\sqrt{n^2+m^2}} \!=\! \frac{(m-n)\ket{+}+(m+n)\ket{-}}{\sqrt{2(n^2+m^2)}}, \label{-l}\\
\!\ket{\widetilde{1}}_B \!=\! \frac{n\ket{0}-m\ket{1}}{\sqrt{n^2+m^2}} \!=\! \frac{(n-m)\ket{+}+(m+n)\ket{-}}{\sqrt{2(n^2+m^2)}}, \label{1l}
\end{eqnarray}
where we have dropped the subindex $B$ at the right hand side of the above expressions for simplicity.
It is not difficult to see that the set 
$\{|\widetilde{0}\rangle_B,|\widetilde{1}\rangle_B\}$
defines an orthonormal basis as well as the set $\{\ket{\widetilde{+}}_B,\ket{\widetilde{-}}_B\}$.
Let us call them the $\tilde{z}$-basis and $\tilde{x}$-basis, respectively.

The analogy with the BB84 protocol is now clear. Whenever Alice teleports the state $\ket{+}$, Bob randomly gets either the state
$|\widetilde{0}\rangle$ or $\ket{\widetilde{+}}$, depending on which Bell measurement she obtained. If the matching
condition is achieved ($n=m$), we have that $|\widetilde{0}\rangle=\ket{\widetilde{+}}=\ket{+}$ and Bob will correctly 
assign the bit value $0$ after measuring his qubit in the x-basis. Similarly, if Alice teleports the 
state $\ket{-}$, Bob's qubit is either $|\widetilde{1}\rangle$ or $\ket{\widetilde{-}}$. For $n=m$ we have
$|\widetilde{1}\rangle=\ket{\widetilde{-}}=\ket{-}$ and Bob will get the correct bit value $1$
after finishing his measurement in the x-basis.

On the other hand, if the matching condition is not fulfilled ($n\neq m$), the probability of Bob making a mistake
is not zero and it is given by $(m-n)^2/(2(n^2+m^2))$. Moreover, the fact that $m\neq n$ breaks the degeneracy of 
Bob's possible states after the teleportation, i.e., $|\widetilde{0}\rangle\neq\ket{\widetilde{+}}$ 
and $|\widetilde{1}\rangle\neq\ket{\widetilde{-}}$.
This induces an encoding into non-orthogonal states for the bit values sent by Alice,  
exactly as it happens in the BB84 protocol. Indeed, in this scenario the
bit value 0 is either associated with $|\widetilde{0}\rangle$ or 
$\ket{\widetilde{+}}$, where $\langle\widetilde{0}|\widetilde{+}\rangle = 2nm/(n^2+m^2)$, and the bit
value 1 is either related to $|\widetilde{1}\rangle$ or 
$\ket{\widetilde{-}}$, where again $\langle\widetilde{1}|\widetilde{-}\rangle = \langle\widetilde{0}|\widetilde{+}\rangle$. 

Putting it another way, although in the GR10 protocol we employ only one set of orthonormal
states to encode the key, the fact that its operation is based on the probabilistic
teleportation protocol leads to an effective non-orthogonal encoding whenever $m\neq n$.
And since Eve cannot always guess the correct values of $m$ and $n$ in her eavesdropping \textit{before} Alice
and Bob finish all their measurements, she will
necessary be caught tampering with the execution of the GR10 protocol. 

Let us now move to the quantitative security analysis of the GR10 protocol. First we note that,
according to the discussion in the previous paragraphs, whenever the matching condition is 
fulfilled ($m=n$) Bob gets an exact replica of the teleported state. Thus, the entanglement-based 
representation of the post-selected successful cases of the GR10 protocol is
\begin{equation}
|\tilde{\Phi}_1\rangle = \frac{1}{\sqrt{2}}(\ket{++}+\ket{--}).
\label{b1+}
\end{equation}
Equation (\ref{b1+}) is the Bell state (\ref{b1}) written in the x-basis.

After Eve's eavesdropping, a possible purification representing the state of Alice, Bob, and Eve is
\begin{equation}
|\Psi \rangle_{ABE} = \sum_{j=1}^{4}\sqrt{\lambda_j}|\tilde{\Phi}_j\rangle_{AB}|\epsilon_j\rangle_E,
\label{psiABE3}
\end{equation}
where $|\tilde{\Phi}_1\rangle$ was just defined and 
\begin{eqnarray}
|\tilde{\Phi}_2\rangle & =& \frac{1}{\sqrt{2}}(|+-\rangle + |-+\rangle),
\label{b2+} \\
|\tilde{\Phi}_3\rangle & =& \ket{01},
\label{b3+} \\
|\tilde{\Phi}_4\rangle & =& \ket{10}.
\label{b4+}
\end{eqnarray}
Equation (\ref{b2+}) is the standard Bell state (\ref{b2}) 
rewritten in the x-basis.

The entanglement-based representation (\ref{psiABE3}) is the one  
we employed when we reassessed the security analysis of the BB84 protocol using a different purification, 
namely, Eq.~(\ref{psiABE2}). 
Here, however,
we deal with the x-basis alone, while in Sec. \ref{newBB84} both the z- and x-bases were
possible preparation and measurement basis for Alice and Bob. The fact that we now have no knowledge 
of $\varepsilon_z$ will reflect in one of the $\lambda$'s being 
undetermined. This forces us to carry out the maximization of Eve's accessible information over all possible strategies that she
might eventually use. 
When compared to the calculations of Sec. \ref{newBB84}, this will lead to a lower value for 
the error rate $\varepsilon_x$ below which we have a positive secret-key fraction and therefore security.

The states teleported by Alice are either $|+\rangle$ or $|-\rangle$. Since they are chosen with
equal chances, Alice's ensemble of qubits are given by $(|+\rangle\langle+|+|-\rangle\langle-|)/2$
or, equivalently, by $(|0\rangle\langle0|+|1\rangle\langle1|)/2$. Hence,
as in Sec. \ref{newBB84}, we must have 
\begin{equation}
p_A(0)=p_A(1)=1/2,
\end{equation}
which leads to the constraint (\ref{constraint}),
\begin{equation}
\lambda_3 = \lambda_4 = \lambda.
\end{equation}

Carrying out similar calculations to the ones detailed in Sec. \ref{bb84}, we get
the following expression for the mutual information between Alice and Bob,
\begin{equation}
I(A:B) = 1 - h(\varepsilon_x).
\end{equation}
The error rate $\varepsilon_x$ is the probability of Bob assigning the wrong
bit value to the bit teleported by Alice and is given by 
\begin{equation}
\varepsilon_x=\lambda_2+\lambda = (1+\lambda_2-\lambda_1)/2,
\label{exgr10}
\end{equation}
where we used the normalization condition
\begin{equation}
\lambda_1+\lambda_2+2\lambda=1 
\label{normgr10}
\end{equation}
to arrive at the last equality.

By the same token we get
\begin{equation}
S(\rho_E) = -\lambda_1\log\lambda_1  -\lambda_2\log\lambda_2 -2\lambda\log\lambda,  
\label{SofE}
\end{equation}
\begin{eqnarray}
\rho_{E|+} &=& \sum_{j=1}^4\lambda_j \ket{\epsilon_j}_{\!E\,E\!}\bra{\epsilon_j} 
+ \sqrt{\lambda_1 \lambda_3/2}\left(\ket{\epsilon_1}_{\!E\,E\!}\bra{\epsilon_3} + h.c.\right) \nonumber \\
&&+ \sqrt{\lambda_1 \lambda_4/2}\left(\ket{\epsilon_1}_{\!E\,E\!}\bra{\epsilon_4} + h.c.\right)  
- \sqrt{\lambda_2 \lambda_3/2}\left(\ket{\epsilon_2}_{\!E\,E\!}\bra{\epsilon_3} + h.c.\right) \nonumber \\
&&+ \sqrt{\lambda_2 \lambda_4/2}\left(\ket{\epsilon_2}_{\!E\,E\!}\bra{\epsilon_4} + h.c.\right),  \\
\label{rhoE|+}
\rho_{E|-} &=& \sum_{j=1}^4\lambda_j \ket{\epsilon_j}_{\!E\,E\!}\bra{\epsilon_j} 
- \sqrt{\lambda_1 \lambda_3/2}\left(\ket{\epsilon_1}_{\!E\,E\!}\bra{\epsilon_3} + h.c.\right) \nonumber \\
&&- \sqrt{\lambda_1 \lambda_4/2}\left(\ket{\epsilon_1}_{\!E\,E\!}\bra{\epsilon_4} + h.c.\right)  
+ \sqrt{\lambda_2 \lambda_3/2}\left(\ket{\epsilon_2}_{\!E\,E\!}\bra{\epsilon_3} + h.c.\right) \nonumber \\
&&- \sqrt{\lambda_2 \lambda_4/2}\left(\ket{\epsilon_2}_{\!E\,E\!}\bra{\epsilon_4} + h.c.\right) . 
\label{rhoE|-}
\end{eqnarray}

The two non-null eigenvalues of either $\rho_{E|+}$ or $\rho_{E|-}$ are $(1+\lambda_1 - \lambda_2)/2=1-\varepsilon_x$ and
$(1+\lambda_2 - \lambda_1)/2=\varepsilon_x$, where we used that $\lambda_3=\lambda_4$ 
and Eq.~(\ref{exgr10}) to write the eigenvalues
as shown. We thus get  
\begin{equation}
S(\rho_{E|+}) = S(\rho_{E|-}) = h(\varepsilon_x)
\end{equation}
and with the aid of Eq.~(\ref{SofE}) we can write the Holevo quantity as follows,
\begin{equation}
\chi(A:E) =  -\lambda_1\log\lambda_1 -\lambda_2\log\lambda_2 -2\lambda\log\lambda - h(\varepsilon_x).
\end{equation}

Finally, the secret-key fraction (\ref{dwb}) becomes
\begin{eqnarray}
r &=& 1 + \min_{\text{Eve}}\left\{\lambda_1\log\lambda_1 +\lambda_2\log\lambda_2 +2\lambda\log\lambda\right\} \nonumber \\
&=& 1 +\min_{\lambda}\left\{(1-\varepsilon_x - \lambda)\log(1-\varepsilon_x - \lambda) 
+(\varepsilon_x - \lambda)\log(\varepsilon_x - \lambda) 
+2\lambda\log\lambda\right\}. \nonumber \\
\label{rgr10}
\end{eqnarray}
To obtain the last equality we used Eqs.~(\ref{exgr10}) and (\ref{normgr10}) to express $\lambda_1$ and
$\lambda_2$ in terms of $\varepsilon_x$ and $\lambda$,
\begin{equation}
\lambda_1 = 1-\varepsilon_x - \lambda, \;\;
\lambda_2 = \varepsilon_x - \lambda. 
\end{equation}

Solving
\begin{equation}
\frac{d r}{d\lambda} = 0 
\end{equation}
for $\lambda$ we get
\begin{equation}
\lambda_{min} = \varepsilon_x (1-\varepsilon_x)
\label{lmin}
\end{equation}
and that 
\begin{equation}
\frac{d^2r(\lambda_{min})}{d\lambda^2}>0, 
\end{equation}
proving that we indeed 
have found the minimum of $r(\lambda)$. 

Inserting Eq.~(\ref{lmin}) into (\ref{rgr10}) we obtain the following lower bound for
the secret-key fraction of the GR10 protocol,
\begin{equation}
r = 1 - 2h(\varepsilon_x).
\label{gr10r}
\end{equation}

\begin{figure}[!ht]
\begin{center}
\includegraphics[width=9cm]{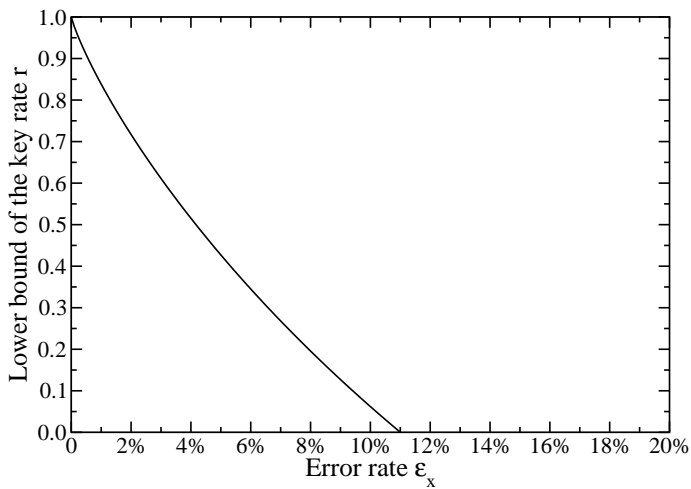}
\caption{
Lower bound of the secret-key fraction $r$ as a function of $\varepsilon_x$ 
for the GR10 protocol (Eq.~(\ref{gr10r})). The quantity $\varepsilon_x$ is the error rate for Bob's measurements 
in the x-basis. Here and in all graphics all quantities are dimensionless.
}
\label{fig1}
\end{center}
\end{figure}

Searching for the root of Eq.~(\ref{gr10r}) we get that $r>0$ if
\begin{equation}
\varepsilon_x \lesssim 11\%.
\end{equation}
In other words, for error rates less than $11\%$ the GR10 protocol operates securely. 
In Fig. \ref{fig1} we show Eq.~(\ref{gr10r}) as a function of $\varepsilon_x$. 

Similarly to what we did when presenting the alternative security proof of the BB84 protocol,
we have worked out the security  
analysis for the GR10 protocol without imposing constraint (\ref{constraint}).
Again, we obtained Eq.~(\ref{constraint}) at the end of the minimization process 
as the optimal scenario for Eve.

Moreover, the purification (\ref{psiABE3}) is crucial to attain a tight lower bound for
the secret-key fraction $r$. Had we worked with the purification (\ref{psiABE}), as usually done in 
the security analysis of the BB84 protocol, we would get a too strong lower bound, one that yields $r<0$ for any value of
$\varepsilon_x>0$. 

Finally, it is important to note that the security analysis presented here and 
in the next section can be directly extended to coherent attacks, leading to the unconditional security
of the GR10 protocol and its modified version given in Sec. \ref{gr10-new}. 
This is true due to the works of  
Renner \cite{ren07} and Renner and Cirac \cite{ren09}, built on the previous works of Refs. \cite{kra05,kra05b}. 
The key result of Ref. \cite{ren07} was the proof that 
in the asymptotic limit security against collective attacks are essentially equivalent to security against 
coherent ones. 

\section{Modifications to the GR10 protocol and their security analysis}
\label{gr10-new}

We now modify the operation of the GR10 protocol in the following sense. Originally, as
given in Ref. \cite{rig10}, all instances in which Alice's measurement results yielded the 
generalized Bell states $|\Phi_1^m\rangle$ and $|\Phi_4^m\rangle$ were discarded as well as the cases where
the matching condition did not occur ($m\neq n$). We now want
to consider a protocol where all cases are included as a valid outcome. In this way we increase 
the size of the raw key $R$ at the cost of decreasing the value of the secret-key fraction $r$. 
Whether the secret key's size $K=rR$ will be greater in one or the other way of implementing the GR10 protocol 
will depend of the value of the error rate and on the values of $n_1$ and $n_2$ adopted in its operation.

When we consider all instances, i.e., all Alice's possible measurement outcomes and whether or not $n=m$, 
Bob will inevitably fail sometimes in the correct identification of the bit value teleported by Alice,
even in the noiseless case. Therefore, the entanglement-based representation of the modified GR10 protocol 
will inherit this property and it is given by
\begin{eqnarray}
\ket{\xi_1} &=& \sqrt{p_{AB}(+,+)}|++\rangle + \sqrt{p_{AB}(-,-)}|--\rangle \nonumber \\
&& + \sqrt{p_{AB}(+,-)}|+-\rangle + \sqrt{p_{AB}(-,+)}|-+\rangle. \nonumber \\
&&
\label{xi}
\end{eqnarray}
Here $p_{AB}(+,+) + p_{AB}(-,-)$ is the probability of Alice and Bob obtaining a perfect correlation between their bit
values and 
$p_{AB}(+,-) + p_{AB}(-,+)$ is the probability of they disagreeing about those values. 

According to the functioning of the probabilistic teleportation protocol, a direct calculation gives
\begin{eqnarray}
p_{AB}(+,+) = p_{AB}(-,-) &=& p, \label{p}\\
p_{AB}(+,-) = p_{AB}(-,+) &=& 1/2-p, \label{1-p}
\end{eqnarray}
where
\begin{equation}
p = \frac{1}{4} + \frac{(n_1+n_2)^2(1+n_1n_2)^2}{4(1+n_1^2)^2(1+n_2^2)^2}.
\label{pn1n2}
\end{equation}
It is worth noting that $p$ is symmetric if we change the value of $n_1$ with that of $n_2$.
Moreover, $p$ is a monotonically increasing function of either $n_1$ or $n_2$ and  
whenever $n_1\neq 0$ or $n_2\neq 0$, $p>1/4$. Also, for $n_1=n_2=1$ we have $p=1/2$.
Since we must have some entanglement shared between Alice and Bob in the operation of the 
GR10 protocol, in what follows $1/4 < p \leq 1/2$.\footnote{Incidentally, it is worth mentioning that if 
in the BB84 protocol we accept all instances as a valid outcome, even when Alice and Bob use different preparation and measurement basis,
we get an error rate of $25\%$ in the ideal case (no Eve or noise). When the matching condition is not satisfied, the outcomes 
of Bob's measurements are completely uncorrelated to the bit values encoded by Alice in the qubits sent to him.
On the other hand, the modified GR10 protocol's ideal error rate, $1-2p$, depends on the entanglement of the quantum states
shared between Alice and Bob (the values of $n_1$ and $n_2$, cf. Eq.~(\ref{pn1n2})).
As such, for the modified GR10 protocol we can tune this error rate as we wish and
whenever $p\geq 3/8\approx 0.375$ the error rate $1-2p$ 
is lower than $25\%$, approaching zero as $n_1$ and $n_2$ tend to one. 
Furthermore, due to the GR10 protocol's teleportation based operation, Bob's measurement outcomes are 
not completely independent of Alice's teleported qubits, 
including the instances in which Alice and Bob assign different bit values
at a given run of the protocol. 
It is this entanglement-dependent ubiquitous correlation that
allows the modified GR10 protocol to operate securely for not too low levels of entanglement.}

Thus, using Eqs.~(\ref{p}) and (\ref{1-p}) the entanglement-based representation (\ref{xi}) becomes
\begin{eqnarray}
\ket{\xi_1} \!&=&\! \sqrt{p}(|++\rangle + |--\rangle) + \sqrt{1/2-p}(|+-\rangle + |-+\rangle) \nonumber \\
&=& \sqrt{2p}|\Phi_1\rangle + \sqrt{1-2p}|\Phi_2\rangle,
\label{xi1}
\end{eqnarray}
where $\ket{\Phi_1}$ and $\ket{\Phi_2}$ are the Bell states (\ref{b1}) and (\ref{b2}).
Note that we have suppressed for simplicity the subscripts $AB$ when writing the above kets. 

The state describing Alice, Bob, and Eve after Eve tampered with the key transmission
can be represented 
by the following purification,
\begin{equation}
|\Psi \rangle_{ABE} = \sum_{j=1}^{4}\sqrt{\lambda_j}|\xi_j\rangle_{AB}|\epsilon_j\rangle_E.
\label{psiABE4}
\end{equation}
Here $|\xi_1\rangle$ is given by Eq.~(\ref{xi1}) and  
\begin{eqnarray}
|\xi_2\rangle & =& \sqrt{2p}|\Phi_2\rangle - \sqrt{1-2p}|\Phi_1\rangle,
\label{xi2} \\
|\xi_3\rangle & =& \ket{\Phi_3},
\label{xi3} \\
|\xi_4\rangle & =& \ket{\Phi_4},
\label{xi4}
\end{eqnarray}
where $|\Phi_3\rangle$ and $\ket{\Phi_4}$ are the Bell states (\ref{b3}) and (\ref{b4}).

As before, this particular way of writing the purification is crucial in our quest for a tight
lower bound for the secret-key fraction $r$. This is true since we will get an additional constraint on 
the possible values of the $\lambda$'s due to the specific operation of the GR10 protocol. 
This constraint is obtained noting that 
since Alice always teleports with equal chances the states $|+\rangle$ or $|-\rangle$,
we must necessarily have,
\begin{equation}
p_A(0)=p_A(1)=1/2.
\label{constraintN}
\end{equation}
Note that $p_A(+)=p_A(-)=1/2$ is trivially satisfied.

A direct calculation gives
\begin{eqnarray}
p_A(0) &=& 1/2 + (\lambda_1 - \lambda_2)\sqrt{2p(1-2p)}, \\
p_A(1) &=& 1/2 - (\lambda_1 - \lambda_2)\sqrt{2p(1-2p)},
\end{eqnarray}
and Eq.~(\ref{constraintN}) is satisfied for $p\neq 1/2$ if, and only if,
\begin{equation}
\lambda_1 = \lambda_2 = \lambda.
\label{constraintNN}
\end{equation}
For $p=1/2$ this restriction is not needed since we automatically have $p_A(0)=p_A(1)=1/2$.
However, employing arguments of continuity for the value of the secret-key fraction as a function of $p$ 
we can set $\lambda_1 = \lambda_2$ for the whole
range of $p$ without getting into any physical 
or mathematical inconsistency.\footnote{We have also employed a different entanglement-based representation
designed to handle the $p=1/2$ case alone. The secret-key fraction
we obtained was the same as the one we got for $p=1/2$ using the present 
entanglement-based representation with the constraint $\lambda_1=\lambda_2$.}

Before we proceed it is important to define in the present context what we mean by 
the ``error'' made by Bob when measuring his qubits due to the presence of Eve. 
Since now even without Eve it is possible that Bob gets the wrong bit value sent by Alice, it is more appropriate to
talk about ``deviation'' from the expected results in the ideal case (no Eve). 
Therefore,
we define the deviation $\delta_x$ as follows,
\begin{equation}
\delta_x \!= \!p_{AB}(+,-) + p_{AB}(-,+) - p^0_{AB}(+,-) - p^0_{AB}(-,+),
\label{deltax}
\end{equation}
where $p^0_{AB}(a,b)$ is the joint probability of Alice and Bob obtaining, respectively,
the values $a$ and $b$ when Eve does not interfere.  Note that
$p^0_{AB}(+,-) + p^0_{AB}(-,+)$ is the probability of Alice and Bob disagreeing about
the bit values in the absence of Eve while $p_{AB}(+,-) + p_{AB}(-,+)$ is the probability of
they disagreeing when Eve is present. 

We also define the relative deviation of agreement as follows,
\begin{eqnarray}
\Delta_x &=&  \frac{p^0_{AB}(+,\!+) + p^0_{AB}(-,\!-) - p_{AB}(+,\!+) - p_{AB}(-,\!-)}{p^0_{AB}(+,\!+) + p^0_{AB}(-,\!-)}, \nonumber \\
& =& \frac{\delta_x}{p^0_{AB}(+,+) + p^0_{AB}(-,-)},
\label{Deltax}
\end{eqnarray}
where the last equality comes from the fact that $\sum_{a,b}p_{AB}(a,b) = 1$ and from Eq.~(\ref{deltax}). 
Note that $\delta_x=\Delta_x=\varepsilon_x$ whenever the probability of making a mistake in the ideal scenario  
is zero, as it happens in the BB84 and in the original GR10 protocols (cf. Eq.~(\ref{errorZ})).

Carrying out the calculations of the joint probabilities we get for Eqs.~(\ref{deltax}) and (\ref{Deltax}),
\begin{eqnarray}
\delta_x &=& \lambda + \lambda_4 - 1 + 2p, \label{deltax2} \\
\Delta_x &=& \delta_x/(2p). \label{Deltax2}
\end{eqnarray}
Using Eqs.~(\ref{constraintNN}) and (\ref{deltax2}), together with the normalization condition
$\sum_{j=1}^4\lambda_j$ $=$ $1$, we can express the four $\lambda$'s as follows,
\begin{eqnarray}
\lambda_1 &=&\lambda_2 = \lambda, \label{eq1e2}\\
\lambda_3 &=& 2p - \delta_x - \lambda = \lambda_+ - \lambda, \label{eq3}\\
\lambda_4 &=& 1-2p + \delta_x -\lambda = \lambda_- - \lambda. \label{eq4}
\end{eqnarray}

This allows us to write the mutual information between Alice and Bob as
\begin{equation}
I(A:B) = 1 - h(2p - \delta_x).
\end{equation}

The relevant quantities needed to the calculation of the Holevo quantity 
are obtained in the same fashion already 
discussed for the BB84 and the original GR10 protocols:
\begin{equation}
S(\rho_E) = -2\lambda\log\lambda  -\lambda_3\log\lambda_3 -\lambda_4\log\lambda_4,  
\label{SofEN}
\end{equation}
\begin{eqnarray}
\rho_{E|+} &=& \sum_{j=1}^4\lambda_j \ket{\epsilon_j}_{\!E\,E\!}\bra{\epsilon_j} 
+ \sqrt{2p\lambda_1 \lambda_3}\left(\ket{\epsilon_1}_{\!E\,E\!}\bra{\epsilon_3} + h.c.\right) \nonumber \\
&&\!\!\!\!\!\!\! - \sqrt{(1-2p)\lambda_1 \lambda_4}\left(\ket{\epsilon_1}_{\!E\,E\!}\bra{\epsilon_4} + h.c.\right)  
- \sqrt{(1-2p)\lambda_2 \lambda_3}\left(\ket{\epsilon_2}_{\!E\,E\!}\bra{\epsilon_3} + h.c.\right) \nonumber \\
&&\!\!\!\!\!\!\!- \sqrt{2p\lambda_2 \lambda_4}\left(\ket{\epsilon_2}_{\!E\,E\!}\bra{\epsilon_4} + h.c.\right),  \\
\label{rhoE|+N}
\rho_{E|-} &=& \sum_{j=1}^4\lambda_j \ket{\epsilon_j}_{\!E\,E\!}\bra{\epsilon_j} 
- \sqrt{2p\lambda_1 \lambda_3}\left(\ket{\epsilon_1}_{\!E\,E\!}\bra{\epsilon_3} + h.c.\right) \nonumber \\
&&\!\!\!\!\!\!\! + \sqrt{(1-2p)\lambda_1 \lambda_4}\left(\ket{\epsilon_1}_{\!E\,E\!}\bra{\epsilon_4} + h.c.\right)  
+ \sqrt{(1-2p)\lambda_2 \lambda_3}\left(\ket{\epsilon_2}_{\!E\,E\!}\bra{\epsilon_3} + h.c.\right) \nonumber \\
&&\!\!\!\!\!\!\!+ \sqrt{2p\lambda_2 \lambda_4}\left(\ket{\epsilon_2}_{\!E\,E\!}\bra{\epsilon_4} + h.c.\right). 
\label{rhoE|-N}
\end{eqnarray}

The eigenvalues of $\rho_{E|+}$ and $\rho_{E|-}$ are the same. The non-zero ones can be written as 
$\lambda_+$ and $\lambda_-$ if we use Eqs.~(\ref{eq1e2})-(\ref{eq4}).
This allows us to write   
\begin{equation}
S(\rho_{E|+}) = S(\rho_{E|-}) = h(\lambda_+) = h(2p - \delta_x).
\end{equation}

The Holevo quantity then becomes 
\begin{equation}
\chi(A:E) =  -2\lambda\log\lambda -(\lambda_+ - \lambda)\log(\lambda_+ - \lambda) -
(\lambda_- - \lambda)\log(\lambda_- - \lambda) - h(\lambda_+),
\end{equation}
leading to the following secret-key fraction
\begin{equation}
r = 1 + \min_{\text{Eve}}\left\{2\lambda\log\lambda +(\lambda_+ - \lambda)\log(\lambda_+ - \lambda) 
+(\lambda_- - \lambda)\log(\lambda_- - \lambda)\right\}.
\label{rgr10mod}
\end{equation}

Remembering that
\begin{equation}
\lambda_+ = 2p-\delta_x,\;\; 
\lambda_- = 1-2p + \delta_x \label{l-}, 
\end{equation}
we can solve 
\begin{equation}
\frac{d r}{d\lambda} = 0 
\end{equation}
for $\lambda$ and get
\begin{equation}
\lambda_{min} = \lambda_+\lambda_- = (2p-\delta_x)(1-2p + \delta_x).
\label{lminN}
\end{equation}
A direct calculation shows that 
\begin{equation}
\frac{d^2r(\lambda_{min})}{d\lambda^2}>0, 
\end{equation}
which proves that we got the minimum value of $r(\lambda)$. 

Therefore, using Eqs.~(\ref{l-}) and (\ref{lminN}),
we can write the lower bound for the secret-key fraction (\ref{rgr10mod}) as follows,
\begin{equation}
r = 1 - 2h(\lambda_+) = 1 - 2h(2p-\delta_x)=1 - 2h[2p(1-\Delta_x)],
\label{gr10rN}
\end{equation}
where the last equality is obtained using Eq.~(\ref{Deltax2}).

\begin{figure}[!ht]
\begin{center}
\includegraphics[width=9cm]{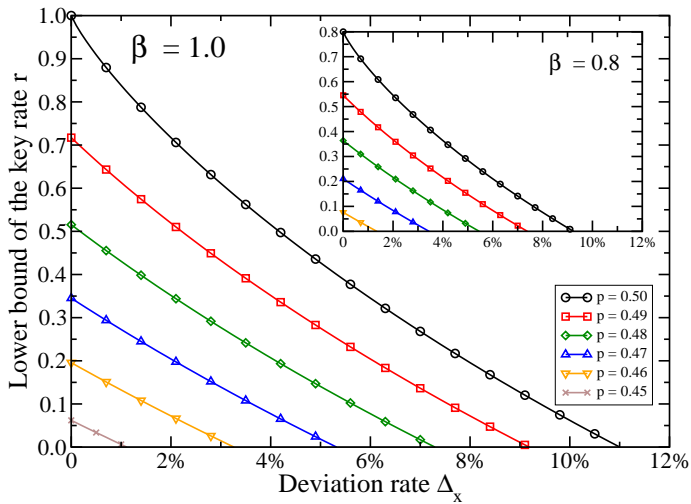}
\caption{
Main panel: Lower bound of $r$, the secret-key fraction for the
modified GR10 protocol as given by Eq.~(\ref{gr10rN}), as a function of $\Delta_x$, the relative deviation from 
the expected probability of Alice and Bob getting the same bit values with and without the presence of Eve. 
Inset: The same as in the main panel but now assuming a reconciliation efficiency of $\beta=0.8$.}
\label{fig2}
\end{center}
\end{figure}

In the main panel of Fig. \ref{fig2} we show $r$ as a function of $\Delta_x$ for several values of
$p$.
We see that the greater the value of $p$ the greater $r$ and thus the greater the secret-key fraction
for a given value of $\Delta_x$. For values of $p$ close to its maximal value $1/2$ we can guarantee security for the 
modified GR10 protocol for a deviation rate as great as $11\%$. As we start decreasing $p$, which is related to 
less entanglement shared between Alice and Bob, we no longer have $r>0$ whenever $p\lesssim 0.45$. In other words, whenever 
the level of entanglement shared between Alice and Bob is such that $p<0.45$, we can only achieve security using the
original GR10 protocol, which works for any value of $p>1/4$ ($n_1\neq 0$ and $n_2\neq 0$). However,
the price to pay as 
we decrease $p$ in the original GR10 protocol is the corresponding reduction of the size of the raw key $R$. 

In the inset of Fig. \ref{fig2} we show how the lower bound to the secret-key fraction responds
to a reconciliation protocol (classical error correction step) whose efficiency is not optimal. 
We model a non-ideal reconciliation protocol by including an effective reduction of the mutual information
between Alice and Bob in the expression for $r$ \cite{gis02,sca09,wee12}. In this scenario it changes to 
\begin{equation}
r = \beta I(A:B) - \max_{\text{Eve}}{\chi(A:E)},
\label{dwbBeta}
\end{equation}
where $0\leq \beta\leq 1$. Repeating all the steps of the previous calculation when we had $\beta = 1$, we get
for the lower bound of the secret-key fraction,
\begin{equation}
r = \beta [1 - h(\lambda_+)] - h(\lambda_+).
\label{dwbBeta2}
\end{equation}

For discrete variable quantum key distribution schemes, the reconciliation protocols 
have $\beta \approx 1$. In any case, we have tested how the modified GR10 protocol responds to 
a reconciliation protocol with $\beta=0.8$, a very conservative value. As we can see in the 
inset of Fig. \ref{fig2}, it still keeps working securely for $p\geq 0.46$ when we have deviation rates
$\Delta_x$ lower or equal to $1\%$ while for $p=0.49$ we have security for $\Delta_x \lesssim 7\%$.

It is worth mentioning that the lower bound for the secret-key fraction computed above implies that 
the modified GR10 protocol works for $n_1=n_2$ and, in particular,
for $n_1=n_2=1$. The latter case corresponds to using the standard and deterministic probabilistic
teleportation protocol to establish a secret-key between Alice and Bob, where Alice teleports only the 
states $|+\rangle$ and $|-\rangle$. In other words, this means that by using the teleportation protocol
as originally presented \cite{ben93}, we can establish a secret key between Alice
and Bob employing only orthogonal states to encode be bit values teleported from Alice to Bob. 
There is no need to use non-orthogonal states
as in the BB84 or B92 protocols \cite{ben84,ben92}. 

This ``counterintuitive'' behavior is readily 
understood by noting that the teleportation protocol has a built-in probabilistic aspect that cannot
be overcome by any ``superpowerful'' quantum Eve. Indeed, if quantum mechanics is the correct description of nature,
no one, not even Eve, has control over what will be the Bell state measured by Alice.
This inherently probabilistic aspect of the quantum teleportation protocol precludes Eve from tampering with
the teleportation protocol without being detected by Alice and Bob. This is so because the unitary operations
that Bob needs to implement on his qubits at the end of the protocol are dictated by what Alice informs him of the results of 
her Bell state measurements and by the entangled resource shared between them. 
In this way, if Eve probes too deep trying to figure out the state teleported to Bob, 
she will cause changes to the Bell state shared between Alice and Bob. As such, the correct unitary operation
that Bob must implement to correct his qubit will change too. Bob, being unaware of this change,
will implement the unitary operation associated with the original Bell state shared with Alice. The use of the wrong unitary
operation will make Bob assign the wrong value for the bit teleported from Alice and will lead
to Alice and Bob detecting the presence of Eve when they compare the bit values of a sample of the shared string of bits. 
Mathematically, this intuitive fact will eventually reflect itself in a positive secret-key rate $r$ even when $n_1=n_2$.\footnote{We can also intuitively understand the security of
the GR10 protocol noting that the states prepared by Alice are not sent from her to Bob, they are teleported, which prevents Eve from having a direct access to those states. 
This is a key difference from the BB84 protocol, where the states encoding the bits are
directly sent to Bob. The teleportation of orthogonal states, instead of their direct 
transmission, together with the inherent random aspect of the measuring results of Alice during the implementation of the teleportation protocol forbid the undetected cloning of those states by Eve.}

We also point to two other ways in which we can modify the GR10 protocol.
The first one corresponds to increasing the number of quantum states employed to encode the bits teleported from Alice
to Bob. We can, similarly to the BB84 protocol, use the states $\ket{0}$ and $\ket{1}$ together  
with the states $\ket{+}$ and $\ket{-}$ to encode the bit values. In this case we will be using
non-orthogonal states to encode the bits and we can think of the GR10 protocol as an additional layer of security 
to the BB84 protocol. Second, we can choose to work with different post-selected 
subensembles. For example, instead of considering all runs of the teleportation protocol as we did above,
we can select only the cases in which Bob sent the partially entangled state with $n=n_1$, discarding the 
$n=n_2$ cases. Or we can work with the cases in which Alice implemented the generalized Bell measurements 
with $m=n_1$, discarding the $m=n_2$ cases. This will reduce the size of the raw key R, but 
if $n_1>n_2$ we can show that by 
working with these subsets we can get a positive secret-key
fraction (security) for a wider range of the deviation $\Delta_x$ when compared to the case where 
all runs of the teleportation protocol is accepted as a valid outcome.\footnote{The entanglement-based representation
and all the calculations leading to the secret-key fraction $r$ when we deal with these subensembles are equal
to the ones shown in Sec. \ref{gr10-new}. The only change is in the value of $p$, Eq.~(\ref{pn1n2}). 
It still depends on $n_1$ and $n_2$ but has a different functional form.}

We end this section by noting that the main technical challenge in the practical 
implementation of the original or of the modified GR10 protocols is the ability to
generate entangled states. Entanglement is
the main ingredient needed to implement the teleportation protocol, which is 
a key building block in the execution of all GR10-like protocols. 
The BB84-like protocols do not need
entangled states to operate. However, once Alice and Bob share an entangled state, 
and it does not need to be a maximally entangled one, they can execute the GR10 protocol no matter how far away they are. This opens the possibility of 
increasing the distance in which a quantum key distribution
scheme works securely, going beyond the maximum distance that any 
BB84-like protocol can reach.

\section{Conclusion}
\label{conclusion}
 
In this work we presented a rigorous security proof for the GR10 quantum key distribution protocol, whose
operation is based on the probabilistic teleportation protocol and on the use of only orthogonal 
quantum states to encode the bits of the secret key \cite{rig10}. 
Being more specific,
we have carried out the asymptotic security analysis of the GR10 protocol against all types of
individual and collective attacks, determining the error rates below which we guarantee a secure
operation of the key distribution scheme. Moreover, applying the results of Refs. \cite{kra05,kra05b,ren07,ren09}
we argued that the present security analysis is easily extended to coherent attacks, leading to the 
unconditional security of the GR10 protocol.

Furthermore, we revisited the security analysis of the BB84 protocol \cite{ben84} by 
exploring the non-uniqueness of the Schmidt decomposition (purification) that describes the 
quantum state of Alice, Bob, and the eavesdropper Eve after the latter has tampered with the key transmission.
This allowed us to show that the BB84 protocol is secure for greater values of error rates than
the standard security analysis predicted \cite{sca09}. This non-uniqueness of the purification
was the key ingredient allowing us to obtain tight lower bounds for the secret-key fraction 
of the GR10 protocol and its modified version.

We also showed a modified version of the GR10 protocol that operates deterministically, 
providing its full security analysis. We showed that 
this version of the GR10 protocol is still secure even if we use only orthogonal states to 
encode the bit values teleported from Alice to Bob. We showed that the price to pay when
going from the probabilistic to the deterministic protocol is a reduction of the value
of the secret-key fraction, specially when the
degree of entanglement shared between Alice and Bob is small.

Finally, we would like to point out two possible extensions of the ideas here presented that we 
believe is worthy of investigation. 
First, it would be interesting to study how to devise a GR10-like protocol that operates with 
qudits instead of qubits. The main quest here is to obtain a scenario where
working with qudits leads to an increase of the error rate below which the protocol is secure.
The second extension is a bit more difficult and it lies in how to
reshape the GR10 protocol in order to make it work with continuous variable systems.

\section*{Acknowledgments}
DL thanks CAPES (Brazilian Agency for the Improvement of Personnel of Higher Education)
for funding and GR thanks the Brazilian agencies CNPq
(National Council for Scientific and Technological Development) and
CNPq and FAPERJ (State of Rio de Janeiro Research Foundation) for financial support through the National Institute of
Science and Technology for Quantum Information.







\begin{thebibliography}{200}

\bibitem{sin00}  Singh, S.: The Code Book: The Science of Secrecy from Ancient Egypt to
Quantum Cryptography. First Anchor Books Edition, New York (2000)

\bibitem{riv79} Rivest, R., Shamir, A., Adleman, L.: On digital signatures and public-key cryptosystems.
MIT Laboratory for Computer Science,
Technical Report, MIT/LCS/TR-212, (1979)

\bibitem{shor97} Shor, P.: Polynomial-Time Algorithms for Prime Factorization and Discrete 
Logarithms on a Quantum Computer. SIAM Journal of Computing \textbf{26}, 1484 (1997)

\bibitem{ben84} Bennett, C.H., Brassard, G.: Quantum cryptography: public key distribution and
coin tossing. In: Proceedings of the IEEE International Conference on
Computers Systems and Signal Processing, Bangalore, India, 175 (1984).

\bibitem{gis02} Gisin, N., Ribordy, G., Tittel, W., Zbinden, H.: Quantum cryptography. Rev. Mod.
Phys. \textbf{74}, 145 (2002)

\bibitem{sca09} Scarani, V., Bechmann-Pasquinucci, H., Cerf,, N.J., Du\v{s}ek, M.,
L\"utkenhaus, N., Peev, M.: The security of practical quantum key distribution. Rev. Mod. Phys. \textbf{81}, 1301 (2009)

\bibitem{wee12}  Weedbrook, Ch., Pirandola, S., Garc\'ia-Patr\'on,  R., 
 Cerf, N.J., Ralph, T.C., Shapiro, J.H., Lloyd, S.: Gaussian quantum information.
Rev. Mod. Phys. \textbf{84}, 621 (2012)

\bibitem{idq} ID Quantique. http://www.idquantique.com. Accessed 11 November 2019

\bibitem{mqt} MagiQ Technologies. http://www.magiqtech.com. Accessed 11 November 2019

\bibitem{ql} Quintessence Labs. https://www.quintessencelabs.com. Accessed 11 November 2019

\bibitem{ben92} Bennett, C.H.: Quantum cryptography using any two nonorthogonal states. Phys. Rev. Lett. \textbf{68}, 3121 (1992)

\bibitem{woo82} Wootters, W.K., Zurek, W.H.: A single quantum cannot be cloned. Nature (London) \textbf{299}, 802 (1982)

\bibitem{die82} Dieks, D.: Communication by EPR devices. Phys. Lett. \textbf{92A}, 271 (1982)

\bibitem{ben93} Bennett,  C.H., Brassard, G., Crepeau, C., Jozsa, R., Peres, A.,
Wootters, W.K.: Teleporting an unknown quantum state via dual classical and Einstein-Podolsky-Rosen channels. Phys. Rev. Lett. \textbf{70}, 1895 (1993)

\bibitem{bri98} Briegel, H.-J., D\"ur, W., Cirac, J.I., Zoller, P.: Quantum Repeaters: The Role of Imperfect Local Operations in Quantum Communication. Phys. Rev.
Lett. \textbf{81}, 5932  (1998)

\bibitem{rig10} Gordon, G., Rigolin, G.: Quantum cryptography using partially entangled states. Opt. Commun. \textbf{283}, 184 (2010)

\bibitem{guo00} Li, W.-L., Li, C.-F., Guo, G.-C.: Probabilistic teleportation and entanglement matching. Phys. Rev. A \textbf{61}, 034301 (2000)

\bibitem{agr02} Agrawal, P., Pati, A.K.: Probabilistic quantum teleportation. Phys. Lett. A \textbf{305}, 12 (2002)

\bibitem{rig06} Gordon, G., Rigolin, G.: Generalized teleportation protocol. Phys. Rev. A \textbf{73}, 042309 (2006)

\bibitem{rig06b} Gordon, G., Rigolin, G.: Generalized quantum-state sharing. Phys. Rev. A \textbf{73}, 062316 (2006)

\bibitem{gor07} Gordon, G., Rigolin, G.: Generalized quantum telecloning. Eur. Phys. J. D \textbf{45}, 347 (2007)

\bibitem{rig09} Rigolin, G.: Unity fidelity multiple teleportation using partially entangled states. J. Phys. B: At. Mol. Opt. Phys. \textbf{42}, 235504 (2009)

\bibitem{for13} Fortes, R., Rigolin, G.: Improving the efficiency of single and multiple teleportation protocols based on the direct use of partially entangled states. Ann. Phys. (N.Y.) \textbf{336}, 517 (2013)

\bibitem{for15} Fortes, R., Rigolin, G.: Fighting noise with noise in realistic quantum teleportation. Phys. Rev. A \textbf{92}, 012338 (2015)

\bibitem{for16} Fortes, R., Rigolin, G.: Probabilistic quantum teleportation in the presence of noise. Phys. Rev. A \textbf{93}, 062330 (2016)

\bibitem{eke91} Ekert, A.K.: Quantum cryptography based on Bell’s theorem. Phys. Rev. Lett. \textbf{67}, 661 (1991)

\bibitem{ben92b} Bennett, C.H., Brassard, G., Mermin, N.D.: Quantum cryptography without Bell’s theorem. Phys. Rev. Lett. \textbf{68}, 557 (1992)

\bibitem{vai95} Goldenberg, L., Vaidman, L.: Quantum Cryptography Based on Orthogonal 
States. Phys. Rev. Lett. \textbf{75}, 1239 (1995)

\bibitem{koa97} Koashi, M., Imoto, N.: Quantum Cryptography Based on Split Transmission of One-Bit Information in Two Steps. Phys. Rev. Lett. \textbf{79}, 2383 (1997)

\bibitem{shu16} Shukla, C., Banerjee, A.,  Pathak, A., Srikanth, R.: Secure quantum communication with orthogonal states. Int. J. Quantum. Inform. \textbf{6}, 1640021 (2016)

\bibitem{sho00} Shor, P.W., Preskill, J.: Simple Proof of Security of the BB84 Quantum Key Distribution Protocol. Phys. Rev. Lett. \textbf{85}, 441 (2000)

\bibitem{lo01} Lo, H.-K.: Proof of unconditional security of six-state quantum key distribution scheme. Quantum Inf. Comput. \textbf{1}, 81 (2001)

\bibitem{dev05} Devetak, I., Winter, A.: Distillation of secret key and entanglement from quantum states. Proc. R. Soc. London, Ser. A \textbf{461}, 207 (2005)

\bibitem{hol73} Holevo, A.S.: Bounds for the Quantity of Information Transmitted by a Quantum Communication Channel. Probl. Inf. Transm. \textbf{9}, 177 (1973)

\bibitem{hug93} Hughston, L.P., Jozsa, R., Wootters, W.K.:  A complete classification of quantum ensembles having a given density matrix. Phys. Lett. A \textbf{183}, 14 (1993)

\bibitem{kra05} Kraus, B., Gisin, N., Renner, R.: Lower and Upper Bounds on the Secret-Key Rate for Quantum Key Distribution Protocols Using One-Way Classical Communication. Phys. Rev. Lett. \textbf{95}, 080501 (2005)

\bibitem{kra05b} Renner, R., Gisin, N., Kraus, B.: Information-theoretic security proof for quantum-key-distribution protocols. Phys. Rev. A \textbf{72}, 012332 (2005)

\bibitem{ren07} Renner, R.: Symmetry of large physical systems implies independence of subsystems. Nat. Phys. \textbf{3}, 645 (2007)

\bibitem{ren09} Renner, R., Cirac, J.I.: de Finetti Representation Theorem for Infinite-Dimensional Quantum Systems and Applications to Quantum Cryptography. Phys. Rev. Lett. \textbf{102}, 110504 (2009)

\bibitem{nie00} Nielsen, M.A., Chuang, I.L.: Quantum Computation and Quantum Information. Cambridge University Press, Cambridge (2000)

\bibitem{ben84b} Bennett, C.H., Brassard, G.,  Bredibart, S., Wiesner, S.: Eavesdrop-detecting quantum communications channel.
IBM Tech. Discl. Bull. \textbf{26}, 4363 (1984) 

\bibitem{bru98} Bru\ss, D.: Optimal Eavesdropping in Quantum Cryptography with Six States. Phys. Rev. Lett. \textbf{81}, 3018 (1998)

\bibitem{bec99} Bechmann-Pasquinucci, H., Gisin, N.: Incoherent and coherent eavesdropping in the six-state protocol of quantum cryptography. Phys. Rev. A \textbf{59}, 4238 (1999)

\end{thebibliography}
\end{document}